\documentclass[10pt,journal,compsoc]{IEEEtran}
\ifCLASSOPTIONcompsoc
  \usepackage[nocompress]{cite}
\else
  \usepackage{cite}
\fi
\ifCLASSINFOpdf
  \usepackage[pdftex]{graphicx}
  \DeclareGraphicsExtensions{.pdf,.jpeg,.png}
\else
\fi
\usepackage{amsmath}
\usepackage{amsfonts}
\DeclareMathOperator*{\argmax}{argmax}
\DeclareMathOperator*{\argmin}{argmin}

\usepackage{algorithm}
\usepackage{algpseudocode}

\algnewcommand\algorithmicinput{\textbf{Input:}}
\algnewcommand\Input{\item[\algorithmicinput]}

\algnewcommand\algorithmicoutput{\textbf{Output:}}
\algnewcommand\Output{\item[\algorithmicoutput]}

\usepackage{booktabs}             
\usepackage{threeparttable}       
\usepackage{siunitx,makecell}     
\usepackage{color,colortbl}       
\definecolor{Gray}{gray}{0.9}

\usepackage{todonotes}

\begin{document}
%

\title{CUDAMPF++: A Proactive Resource Exhaustion Scheme for Accelerating Homologous Sequence Search on CUDA-enabled GPU}

%
%
%
%

\author{Hanyu~Jiang,~\IEEEmembership{Student Member,~IEEE,}
        Narayan~Ganesan,~\IEEEmembership{Senior Member,~IEEE,}
        and~Yu-Dong~Yao,~\IEEEmembership{Fellow,~IEEE}
\IEEEcompsocitemizethanks{\IEEEcompsocthanksitem H. Jiang, N. Ganesan, Y. Yao
are with the Department of Electrical and Computer Engineering, Stevens
Institute of Technology, Hoboken, NJ
07030.\protect\\
E-mail: \{hjiang5, nganesan, Yu-Dong.Yao\}@stevens.edu.}
\thanks{Manuscript received xxxx xx, xxxx; revised xxxx xx, xxxx.}}

%
%

\markboth{Journal of \LaTeX\ Class Files,~Vol.~14, No.~8, August~2015}%
{Shell \MakeLowercase{\textit{et al.}}: Bare Demo of IEEEtran.cls for Computer Society Journals}
%



\IEEEtitleabstractindextext{%
\begin{abstract}
Genomic sequence alignment is an important research topic in bioinformatics and continues to attract significant efforts. As genomic data grow exponentially, however, most of alignment methods face challenges due to their huge computational costs. HMMER, a suite of bioinformatics tools, is widely used for the analysis of homologous protein and nucleotide sequences with high sensitivity, based on profile hidden Markov models (HMMs). Its latest version, HMMER3, introdues a heuristic pipeline to accelerate the alignment process, which is carried out on central processing units (CPUs) with the support of streaming SIMD extensions (SSE) instructions. Few acceleration results have since been reported based on HMMER3. In this paper, we propose a five-tiered parallel framework, CUDAMPF++, to accelerate the most computationally intensive stages of HMMER3's pipeline, multiple/single segment Viterbi (MSV/SSV), on a single graphics processing unit (GPU). As an architecture-aware design, the proposed framework aims to fully utilize hardware resources via exploiting finer-grained parallelism (multi-sequence alignment) compared with its predecessor (CUDAMPF). In addition, we propose a novel method that proactively sacrifices L1 Cache Hit Ratio (CHR) to get improved performance and scalability in return. A comprehensive evaluation shows that the proposed framework outperfroms all existig work and exhibits good consistency in performance regardless of the variation of query models or protein sequence datasets. For MSV (SSV) kernels, the peak performance of the CUDAMPF++ is 283.9 (471.7) GCUPS on a single K40 GPU, and impressive speedups ranging from 1.x (1.7x) to 168.3x (160.7x) are achieved over the CPU-based implementation (16 cores, 32 threads).
\end{abstract}

}

\maketitle

\IEEEdisplaynontitleabstractindextext

%
\IEEEpeerreviewmaketitle


%
%
%
%

\IEEEraisesectionheading{\section{Introduction}\label{sec:introduction}}
\IEEEPARstart{T}{ypical} algorithms and applications in bioinformatics, computational biology and system biology share a common trait that they are computationally challenging and demand more computing power due to the rapid growth of genomic data and the need for high fidelity simulations. As one of the most important branches, the genomic sequence analysis with various alignment methods scales the abstraction level from atoms to RNA/DNA molecules and even whole genomes, which aims to interpret the similarity and detect homologous domains amongst sequences~\cite{Nobile2016Graphics}. For example, the protein motif detection is key to identify conserved protein domains within a known family of proteins. This paper addresses HMMER~\cite{Eddy1998Profile, Eddy2011Accelerated}, a widely used toolset designed for the analysis of homologous protein and nucleotide sequences with high sensitivity, which is carried out on central processing units (CPUs) originally.

HMMER is built on the basis of probabilistic inference methods with profile hidden Markov models (HMMs)~\cite{Eddy2011Accelerated}. Particularly, the profile HMM used in HMMER is Plan-7 architecture that consists of five main states (\textit{Match}(\textit{M}), \textit{Insert}(\textit{I}), \textit{Delete}(\textit{D}), \textit{Begin}(\textit{B}) and \textit{End}(\textit{E})) as well as five special states (\textit{N}, \textit{C}, \textit{J}, \textit{S} and \textit{T}). The \textit{M}, \textit{I} and \textit{D} states which are in the same position form a node, and the number of nodes included in a profile HMM indicates its length. The digital number ``7" in Plan-7 refers to the total of seven transitions per node, which exist in the architecture and each has a transition probability. In addtion, some states also have emission probabilities. This architecute is a little bit different from the original one proposed by Krogh~\textit{et al.}~\cite{Krogh1994Hidden} which contains extra $I$-$D$ and $D$-$I$ transitions.

The profile HMMs employ position-specific \textit{Insert} or \textit{Delete} probabilities rather than gap penalties, which enables HMMER to outperform BLAST~\cite{Altschul1990Basic} on senstivity~\cite{Eddy2011Accelerated}. However, the previous version of HMMER, HMMER2, suffers the computational expense and gains less utilization than BLAST. Due to well-designed heuristics, BLAST is in the order of 100x to 1000x faster than HMMER2~\cite{Eddy2011Accelerated}. Therefore, numerous acceleration efforts have been made for HMMER2, such as~\cite{Walters2009Evaluating, Ganesan2010Accelerating, maddimsetty2006Accelerator, oliver2008integrating, takagi2009accelerating}. Most of them employ application accelerators and co-processors, like field-programmable gate array (FPGA), graphics processing unit (GPU) and other parallel infrastructures, which provide good performence improvement. To popularize HMMER for standard commodity processors, Eddy~\textit{et al.} propose new versions of HMMER, HMMER3 (v3.0) and its subsequent version (v3.1), which achieve the comparable performance as BLAST~\cite{Eddy2011Accelerated}. As the main contribution, HMMER3 implements a heuristic pipeline in \textit{hmmsearch} which aligns a query model with the whole sequence dataset to find out significantly similar sequence matches. The heuristic acceleration pipeline is highly optimized on CPU-based systems with the support of streaming SIMD extensions (SSE) instructions, and hence only few acceleration attempts, including~\cite{Abbas2010Accelerating, li2010A, Lin2014Implementing, Lin2015Accelerating, Ferreira2014Cache, Neto2015Acceleration} and~\cite{Jiang2015Fine}, report further speedups.

Our previous work~\cite{Jiang2016CUDAMPF}, CUDAMPF, proposes a multi-tiered parallel framework to accelerate HMMER3 pipeline on a single GPU, which was shown to exceed the current state-of-the-art. However, the performance evaluation shows that the thoughput of the computational kernel depends on the model length, especially for small models, which implies underutilization of the GPU. This inspires us to exploit finer-grained parallelism, compared with the framework presented in~\cite{Jiang2016CUDAMPF}. In this paper, we describe another tier of parallelization that aims to fully take advantage of the hardware resources provided by single GPU. A novel optimization strategy that proactively utilizes on-chip cache system is proposed to further boosts kernel throughput and improves scalability of the framework. A comprehensive evaluation indicates that our method exhibits good consistency of performance regardless of query models and protein sequence datasets. The generalization of the proposed framework as well as performance-oriented suggestions are also discussed.

The rest of the paper is organized as follows. Section 2 presents background of HMMER3 pipeline, GPU architecture and highlighted CUDA features, followed by a review of CUDAMPF implementation. In Section 3, the in-depth description of our proposed framework is presented. Then, we give comprehensive evaluations and analyses in Section 4. Related works and discussions are presented in Section 5 and 6, respectively. Finally, we present the conclusion of this paper.


\section{Background}
In this section, we go through the new heuristic pipeline of HMMER3 and highlights its computationally intensive stages. An overview of the GPU architecture and the CUDA programming model is also presented. For better understanding of subsequent ideas, we briefly review our previous work, CUDAMPF, at end of this section.

\subsection{Heuristic Pipeline in HMMER3}
\begin{figure}[!h]
\includegraphics[width=0.48\textwidth]{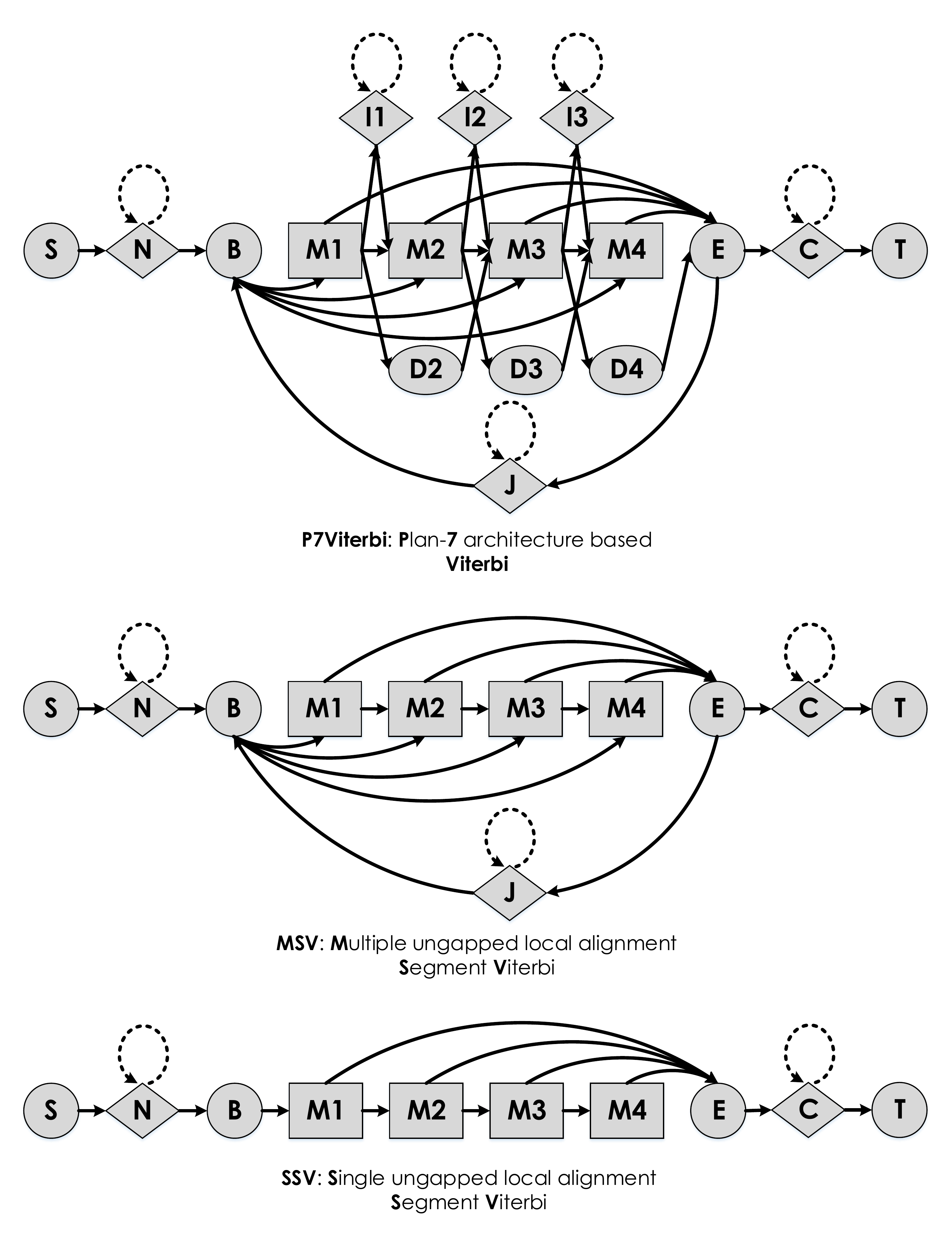}
\centering
\caption{Profiles of P7Viterbi, MSV and SSV models.}
\label{fig:p7profile}
\end{figure}

The main contribution that accelerates the HMMER3 is a new algorithm, multiple segment Viterbi (MSV)~\cite{Eddy2011Accelerated}, which is derived from the standard Viterbi algoritm. The MSV model is a kind of ungapped local alignment model with multiple hits, as shown in Fig.~\ref{fig:p7profile}, and it is achieved by pruning \textit{Delete} and \textit{Insert} states as well as their transitions in the original profile HMMs. The $M$-$M$ transitions are also treated as constants of 1. In addition to the MSV algorithm, another simpler algorithm, single segment Viterbi (SSV), is also introduced to boost the overall performance further. Given that the $J$ state is the bridge between two matched alignments, the SSV model assumes that there is rarely a matched alignment with a score that is higher than the cost of going through the $J$ state, and hence it speculatively removes the $J$ state to gain a significant speedup~\cite{Bjarne2015ssvfilter}. However, in order to avoid false negatives, the SSV model is followed by regular MSV processing to re-calculate suspected sequences. Fig.~\ref{fig:p7profile} illustrates profiles of P7Viterbi, MSV and SSV models with an example of 4 nodes. The solid arrows indicate transitions between different types of states whereas dashed arrows represent the self-increase of a state.

In the pipeline, SSV and MSV models work as heuristic filters (stages) that filter out nonhomologous sequences. All sequences are scored during SSV and MSV stages, and only about 2.2\% of sequences are passed to the next stage, given a threshold. The second stage consists of the P7Viterbi model which only allows roughly 0.1\% of sequences pass, and resulting sequences are then scored with the the full Forward algorithm~\cite{Eddy2011Accelerated}. These four stages mentioned above form the main part of HMMER3's pipeline. However, SSV and MSV stages consumes more than 70\% of the overall execution time~\cite{Jiang2016CUDAMPF,Jiang2015Fine}, and hence they are prime targets of optimization.

\begin{figure}[!h]
\includegraphics[width=0.48\textwidth]{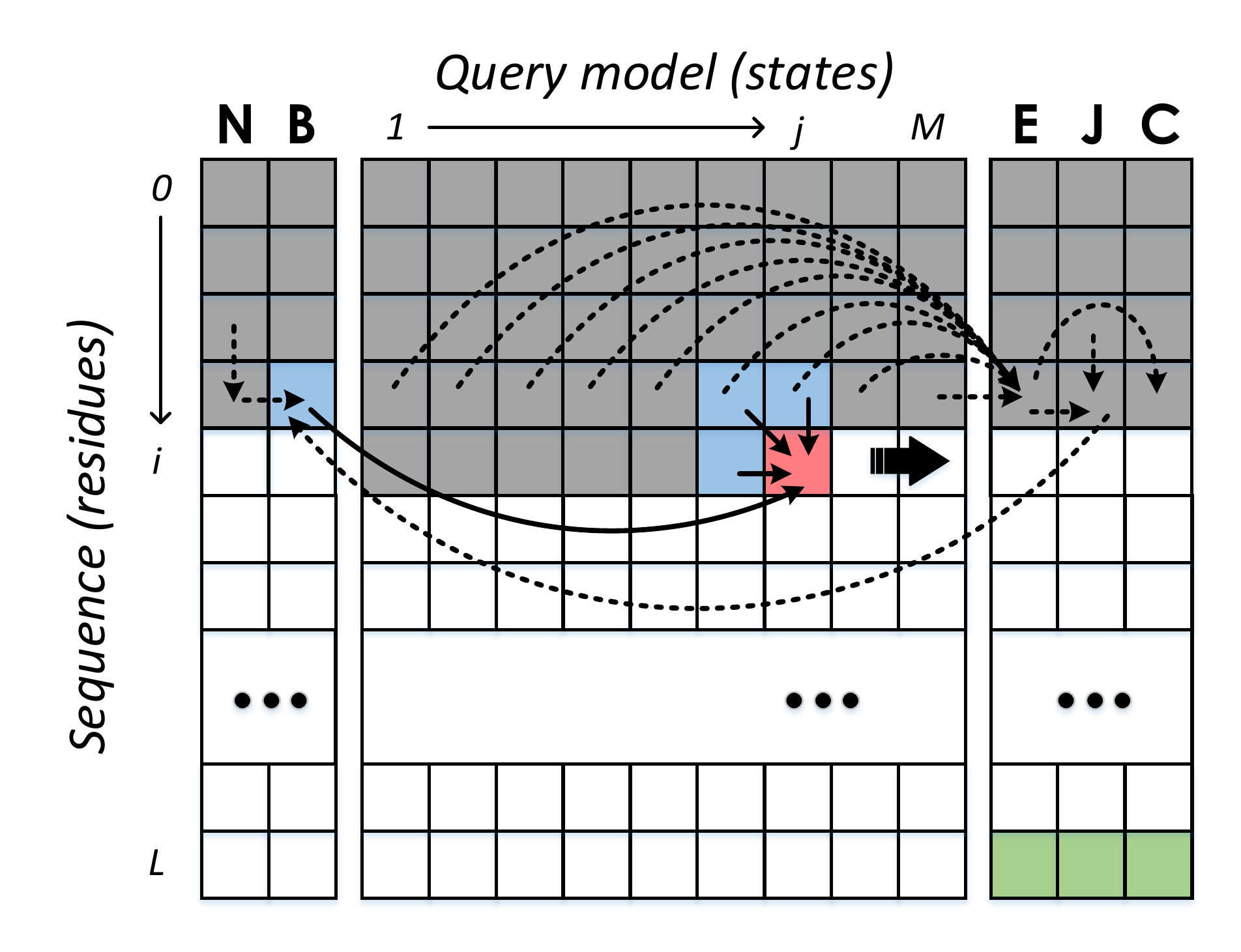}
\centering
\caption{Dynamic programming matrix of the P7Viterbi stage.}
\label{fig:dp}
\end{figure}

Fig.~\ref{fig:dp} illustrates the dynamic programming (DP) matrix of the P7Viterbi stage, corresponding to Fig.~\ref{fig:p7profile}. A lattice of the middle region contains three scores for \textit{Match}, \textit{Insert} and \textit{Delete} states, respectively, whereas flanking lattices only have one. To complete the alignment of a sequence, we need to calculate every lattice of the DP matrix starting from the left-top corner to the right-bottom corner (green) in a row-by-row order, which is a computationally intensive process. In the middle region of the DP matix, each lattice (red) depends on four cells (blue) directly, denoted by solid arrows, which can be formulated as:

\begin{equation}\label{midscores}
\begin{split}
&V_M[i,j]=\epsilon_{M}+\max\begin{cases}
               V_M[i-1,j-1]+T(M_{j-1},M_{j}),\\
               V_I[i-1,j-1]+T(I_{j-1},M_{j}),\\
               V_D[i-1,j-1]+T(D_{j-1},M_{j}),\\
               B[i-1]+T(B,M_{j})
          \end{cases}\\
&V_I[i,j]=\epsilon_{I}+\max\begin{cases}
               V_M[i-1,j]+T(M_{j},I_{j}),\\
               V_I[i-1,j]+T(I_{j},I_{j})\\
          \end{cases}\\
&V_D[i,j]=\max\begin{cases}
               V_M[i,j-1]+T(M_{j-1},D_{j}),\\
               V_D[i,j-1]+T(D_{j-1},D_{j})\\
          \end{cases}
\end{split}
\end{equation}
where $\epsilon$ denotes emission scores. $V$ and $T$ represent scores of $M$/$I$/$D$ states and transitions, respectively. As for MSV and SSV stages, the mathematical formula can be simplified via removing $V_I$ and $V_D$, which results in moderate dependencies and fewer amount of computation than the P7Viterbi stage. However, in order to exceed the performance of the highly optimized CPU-based implementation with the SIMD vector parallelization, it is imperative to go beyond general methods and exploit more parallelism on other multi/many-core processor architectures.

\subsection{GPU Architecture and CUDA Programming Model}
As parallel computing engines, CUDA-enabled GPUs are built around a scalable array of multi-threaded streaming multiprocessors (SMs) for large-scale data and task parallelism, which are capable of executing thousands of threads in the \textit{single-instruction multiple-thread} (SIMT) pattern~\cite{cudaprogrammingguide}. Each generation of GPU introduces more hardware resources and new features, which aims to deal with the ever-increasing demand for computing power in both industry and academia. In this paper, we implement our design on Tesla K40 GPU of Kepler GK110 architecture which equips with 15 powerful streaming multiprocessors, also known as SMXs. Each SMX consists of 192 single-precision CUDA cores, 64 double-precision units, 32 special function units and load/store units~\cite{GK110210whitepaper}. The architecture offers another 48KB on-chip read-only texture cache with an independent datapath from the existing L1 and shared memory datapath, and the maximum amount of available registers for each thread is increased to 255 instead of prior 63 per thread. Moreover, a set of Shuffle instructions that enables a warp of threads to share data without going through shared memory are also introduced in Kepler architecture. This new feature is heavily used in our proposed framework.

The CUDA programming model is designed for NVIDIA GPUs, and it provides users with a development environment to easily leverage horsepower of GPUs. In CUDA, a \textit{kernel} is usually defined as a function that is executed by all CUDA \textit{threads} concurrently. Both \textit{grid} and \textit{block} are vitural units that form a thread hierarchy with some restrictions. Although CUDA allows users to launch thousands of threads, only a warp of threads (32 threads, currently) guarantee that they advance exectuions in lockstep, which is scheduled by a warp scheduler. Hence, the full efficiency is achieved only if all threads within a warp have the same execution path. Traditionally, in order to make sure that threads keep the same pace, a barrier synchronization has to be called explicitly, which imposes additional overhead.

\subsection{CUDAMPF}
In~\cite{Jiang2016CUDAMPF}, we proposed a four-tiered parallel framework, CUDAMPF, implemented on single GPU to accelerate SSV, MSV and P7Viterbi stages of \textit{hmmsearch} pipeline. The framework describes a hierarchical method that parallelizes algorithms and distributes the computational workload considering available hardware resources. CUDAMPF is completely \textit{warp-based} that regards each resident warp as a compute unit to handle the exclusive workload, and hence the explict thread-synchronization is eliminated. Instead, the built-in warp-synchronism is fully utilized. A warp of threads make the alignment of one protein sequence one time and then pick up next scheduled sequence. Given that 8-bit or 16-bit values are sufficient to the precision of algorithms, we couple SIMT execution mechanism with SIMD video instructions to achieve 64 and 128-fold parallelism within each warp. In addition, the runtime compilation (NVRTC), first appeared in CUDA v7.0, was also incorporated into the framework, which enabled swichable kernels and innermost loop unrolling to boost the performance further. CUDAMPF yields upto 440, 277 and 14.3 GCUPS (giga cells updates per second) with strong scalability for SSV, MSV and P7Viterbi kernels, respectively, and it has been proved to exceed all existing work.

\section{Proposed Framework: CUDAMPF++}
This section presents detailed implementations of the proposed framework, CUDAMPF++, that is designed to gain more parallelism based on CUDAMPF. We first introduce a new tier of parallelism followed by a data reformatting scheme for protein sequence data, and then in-depth explanations of kernel design are presented. Finally, we discuss the optimizations of the proposed framework.

\subsection{Five-tiered Parallelism}
In CUDAMPF, the four-tiered parallel framework is proposed to implement MSV, SSV and P7Viterbi kernels. Although the performance improvement is observed on all accelerated kernels, the speedup on P7Viterbi kernel is very limited whereas MSV/SSV kernel yields significant improvement. Given the profiling information~\cite{Jiang2016CUDAMPF}, we are able to gain additional insights into the behaviors: (a) L1 Cache Hit Ratio (CHR) of the P7Viterbi kernel degrades rapidly as model size increases, and (b) its register usage always exceed the maximum pre-allocation for each thread, which indicates the exhaustion of on-chip memory resources and serious register spill. As for MSV/SSV kernels, however, the on-chip memory resources are sufficient. A large amount of low-latency registers, especially when aligning with small models, are underutilized. This can also be proved by performance curves in~\cite{Jiang2016CUDAMPF} in which only upward slopes are observed without any flat or downward trends as model size increases from 100 to 2405. The underutilization leaves an opportunity to exploit further parallelism that can fully take advantage of hardware resources on GPUs.

\begin{figure}[!t]
\includegraphics[width=0.48\textwidth]{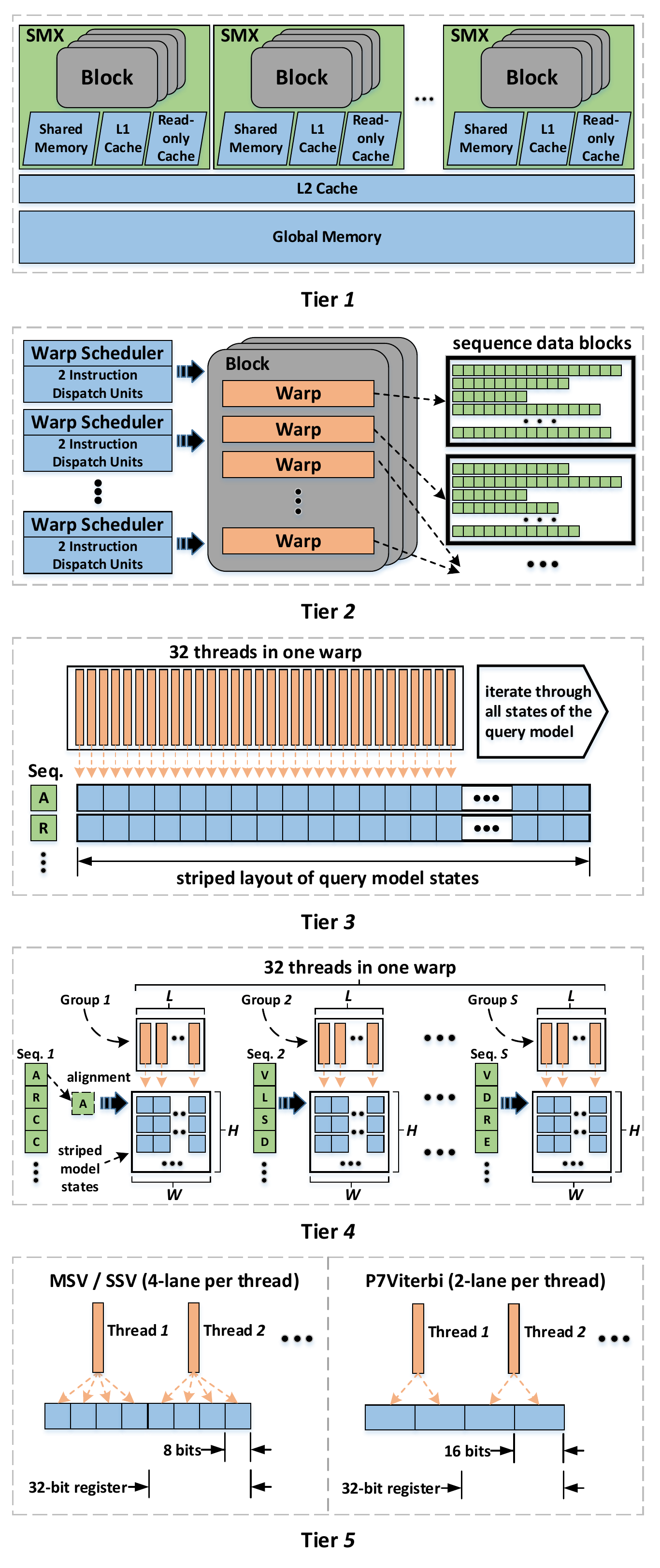}
\centering
\caption{Five-tiered parallel framework based on the NVIDIA Kepler architecture.}
\label{fivetier}
\end{figure}

In addition to original four-tiered structure, another tier of parallelism, inserted between 3rd and 4th tiers, is proposed to enable each warp handle multiple alignments with different protein sequences in parallel while the design of the CUDAMPF only allow single-sequence aligment per warp. This scheme aims to exhaust on-chip memory resources, regardless of the model length, to maximize the throughput of MSV/SSV kernels. Fig.~\ref{fivetier} illustrates the five-tiered parallel framework.

The first tier is based on multiple SMXs that possess plenty of computing and memory resources individually. Given the inexistence of data and execution dependency between each protein sequence, it is straightforward to partition whole sequence database into several chunks and distribute them to SMXs, as the most basic data parallelism. Tier 2 describes the parallelism between multiple warps that reside on each SMX. Since our implementation still applies the warp-synchronous execution that all warps are assigned to process different sequences without inter-warp interactions, explicit synchronizations are eliminated completely. Unlike the CUDAMPF in which warps move to their next scheduled task once the sequence at hand is done, the current design allocates a sequence data block to each warp in advance. A data block may contain thousands of sequences, less or more, depending on the total size of protein sequence dataset and available warps. For multi-sequence alignment, all sequences within each data block need to be reformatted as the striped layout, which enables coalesced access to the global memory. Details of data reformatting will be discussed in Sec.~\ref{dataformat}. The number of resident warps per SMX is still fixed to 32 due to the complexity of MSV and SSV algorithms, which ensures that every thread obtains enough registers to handle complex execution dependencies and avoids excessive register spill. Besides, each SMX contains 4 warp schedulers, each with dual instruction dispath units~\cite{GK110210whitepaper}, and hence it is able to dispatch upto 8 independent instructions each cycle. Those hardware resources have the critical influence on the parallelism of tier 2 and the performance of warp-based CUDA kernels.

Tier 3 is built on the basis of warps. A warp of threads update different model states simultaneously and iterate over remaining model states in batches. The number of iterations depends on the query model size. Once the alignment of an amino-acid residue is done, such as the 'A' and its alignment scores (marked as a row of blue lattices) shown in Fig.~\ref{fivetier} (Tier 3), the warp moves to next residue 'R' and start over the alignment until the end of current sequence. On the basis of tier 3, tier 4 illustrates the multi-sequence alignment that a warp of threads are evenly partitioned into several groups, each has $L$ threads, to make alignments with $S$ different sequences in parallel. For example, the first residues of $S$ sequences, like 'A' of Seq. 1, 'V' of Seq. 2 and 'V' of Seq. S, are extracted together for the first-round alignment. Each group of threads update $W$ scores per iteration, and $H$ iterations are required to finish one alignment. The model states are formatted as a rectangle with $W \times H$ lattices. Considering two models, a large model $M_{l}$ and a small model $M_{s}$ where the size of $M_{l}$ is $S$ times larger than the size of $M_{s}$, we are able to get $W_{s}=\frac{1}{S}W_{l}$ given $H_{s}=H_{l}$, which provides $S$-lane parallelism and roughly keeps register utilization of $M_{s}$ as same as $M_{l}$. The tier 5 remains unchanged as the fine-grained data parallelism: every thread can operate on four 8-bit values and two 16-bit values simutaneously, using single SIMD video instruction~\cite{ptxisa5}, for MSV/SSV and P7Viterbi algorithms, respectively. With the support of tier 5, the parallelism of tier 4 is further extended because each thread takes charge of four different sequences at most. The value of $S$, as the number of sequences processed in parallel by a warp, is defined as below:
\begin{equation}\label{svaluerange}
\lbrace S\mid S=2^{i},i\in\mathbb{Z}\cap[1,\log_2{\frac{\hat{s}\hat{w}_{r}}{\hat{w}_{v}}}] \rbrace,
\end{equation}
where $\hat{s}$ is the warp size, $\hat{w}_{r}$ and $\hat{w}_{v}$ represent the width of registers and participant values, respectively. With Eq.~\ref{svaluerange}, the rest of values can be also formulated as:
\begin{equation}\label{othervalue}
\begin{split}
&W=\frac{\hat{s}\hat{w}_{r}}{\hat{w}_{v}S},
\\
&L=\lceil \frac{\hat{s}}{S}\rceil\,
\\
&H=\max\{2,\lceil \frac{\hat{m}}{W}\rceil\},
\end{split}
\end{equation}
where $\hat{m}$ represents the size of query model. $W$, $L$ and $H$ can be regarded as functions of $S$.

\subsection{Warp-based Sequence Data Blocks}
\label{dataformat}
Due to the introduction of the multi-sequence alignment, loading sequence data in the sequential layout is highly inefficient. Originally, in~\cite{Jiang2016CUDAMPF}, residues of each sequence are stored in contiguous memory space, and warps always read 128 residues of one sequence by one coalesced global memory transaction. As for current design, however, the sequential data layout may lead to 128 transactions per memory request while extracting residues from 128 different sequences. Hence, a striped data layout is the most straightforward solution. Given that warps have exclusive tasks, we propose a data reformatting method that arranges sequences in a striped layout to (a) achieve fully coalesced memory access and (b) balance workload amongst warps. The proposed method partitions large sequence dataset into blocks based on the number of resident warps.

\begin{figure}[!t]
\includegraphics[width=0.48\textwidth]{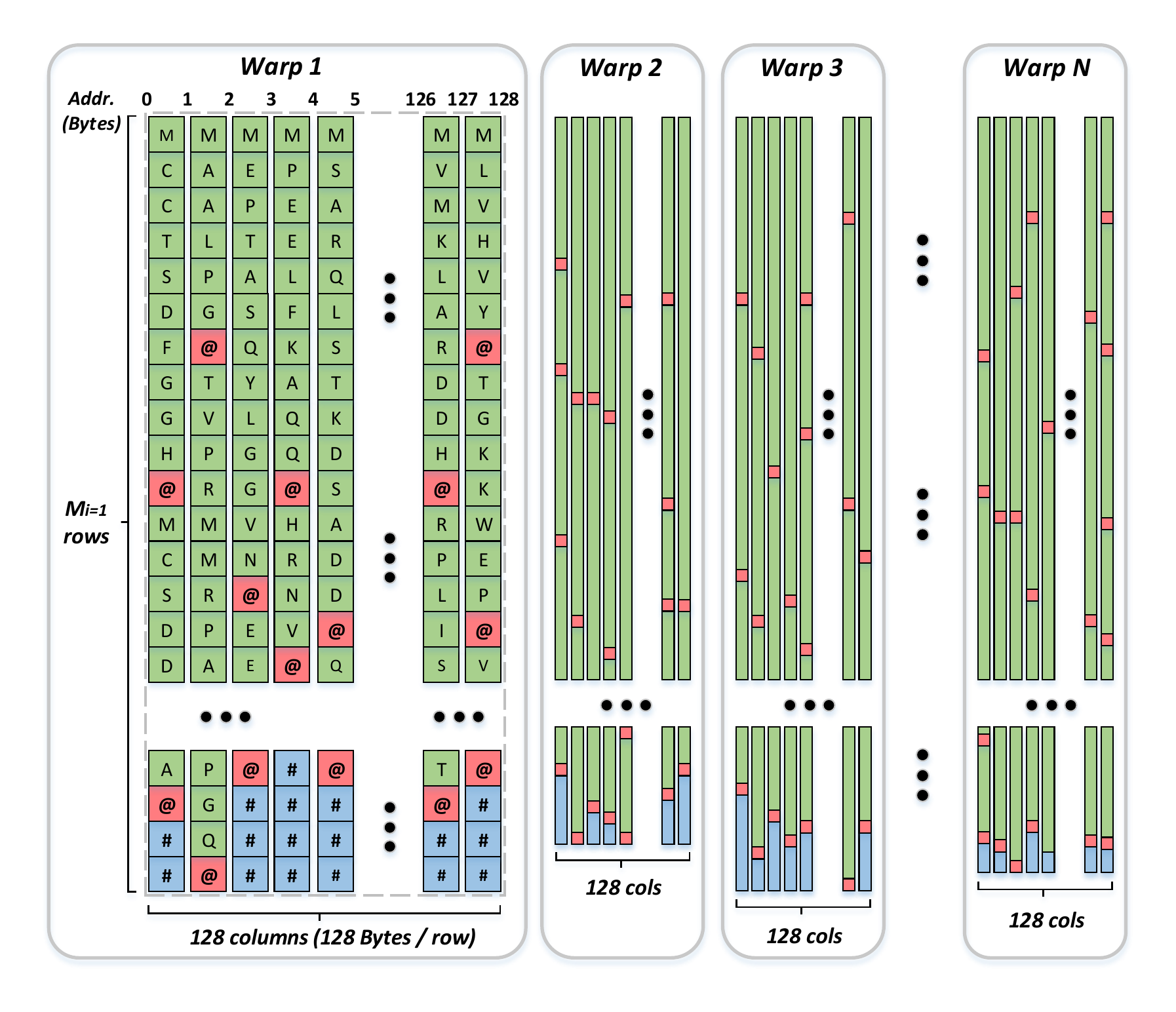}
\centering
\caption{Warp-based sequence data blocks.}
\label{dataformatfigure}
\end{figure}

As shown in Fig.~\ref{dataformatfigure}, all protein sequences are divided into $N$ blocks, each consists of $M_i\times 128$ residues, where $N$ is the total number of resident warps, and $M_i$ represents the height of each block with $i=[1,2,3...N]$. The number $128$, as the width of each block, is pre-fixed for two reasons: (a) MSV/SSV kernels only need participant values with the width of $\hat{w}_{v}=8$ bits, and hence a warp can handle up to $S=128$ sequences simultaneously. (b) 128 residues, each occupies 1 byte, achieves aligned memory address for coalescing access. Marked as different colors, residues consist of three types: regular (green), padding (blue) and ending (red). In each block, sequences are concatenated and filled up 128 columns. A \textit{ending} residue '@' is inserted at the end of each sequence, which is used to trigger an intra-wrap branch to calculate and record the score of current sequence in the kernel. The value of $M_i$ is equal to the length of longest column within each block, such as second column of the data block for warp 1. As for the rest of columns whose length is less than $M_i$, \textit{padding} residue '\#'s are attached to make them all aligned. Residues that belong to the same warp are stored together in row-major order.

\begin{algorithm}
\caption{Protein sequence data reformatting}\label{dataformatalg}
\begin{algorithmic}[1]
\Input all sequences sorted by length in descending order $Seq$, and some parameters, such as $S$.
\Output warp-based sequence data blocks $B_i$.
\State $N\gets N_{smx}*N_{warp}$
\State $lines\gets N*S$
\State create $lines$ containers $C_x$ where $x\in\mathbb{Z}\cap[0,lines)$.
\State $C_x\gets Seq[x]$\Comment{load first $lines$ sequences}
\State $ptr\gets lines$\Comment{start from last container}
\State $maxLen\gets\max\{C_x.len()\}$
\ForAll{$Seq[y]$ where $y\geq lines$}
\Repeat
\State $ptr\gets ptr-1$
\If{$ptr<0$}\Comment{position is not found}
\State $ptr\gets lines$\Comment{start over}
\State $C_{(ptr-1)}.attach(Seq[y])$
\ElsIf{$Seq[y].len()+C_{ptr}.len()\leq maxLen$}
\State $C_{ptr}.attach(Seq[y])$
\EndIf
\Until{$Seq[y]$ is attached.}
\State $maxLen.checkMax()$
\State $ptr\gets lines$ \textbf{if} $ptr<0$.\Comment{refresh pointer normally}
\EndFor\Comment{all sequences are attached}
\State $M_i\gets C_x.divide(N).getMax(S)$ where $i=[1,...,N]$
\State $B_i\gets C_x.padding(M_i)$
\State \textbf{return} $B_i$
\end{algorithmic}
\end{algorithm}

Algorithm~\ref{dataformatalg} shows the pseduo-code of reformatting protein sequence data. $N$ is determined by the number of SMXs and the number of resident warps per SMX, denoted by $N_{smx}$ and $N_{warp}$, respectively (line 1). Given $S$, $N*S$ containers $C_x$ are created to hold and shape sequences as we need. One container corresponds to one column as shown in Fig.~\ref{dataformatfigure}. We load first $N*S$ seqeunces as the basis (line 4), and the \textbf{for} loop (line 7 to 19) iterates over the rest to distribute them into $C_x$. To avoid serious imbalance, the $maxLen$ (line 6) is employed to monitor the longest container as the upper limit. Every new sequence searches for a suitable position based on the accumulated length of each container and $maxLen$ (line 13). We force the sequence to be attached to the last container (line 12) if no position is found. $maxLen$ always be checked and updated by $checkMax()$ function once a new sequence is attached successfully (line 17). $C_x$ are evenly divdied into $N$ blocks when all sequences are attached, and the length of longest container within each block $M_i$ is recorded by $getMax(S)$ function (line 20). The padding process, as the last step (line 21), shapes warp-based sequence data blocks $B_i$ into rectangles. Implementation of this method and the evaluation of workload balancing are presented in Sec.~\ref{workloadeval}.

\subsection{Kernel Design}
\label{kerneldesign}
Since the number of sequences that are processed in parallel is within the range of $\{2,4,8,16,32,64,128\}$, the proposed algorithms are designed to cover all these cases. Therefore, 14 different types of kernels are generated, named as $S$-lane MSV/SSV kernels, and their implementations slightly vary with value $S$.

Algorithm~\ref{MSVkernel} outlines the $S$-lane MSV kernels that is more complex than the implementation of single-sequence alignment in~\cite{Jiang2016CUDAMPF}. Some features are inherited, like (a) using local memory to hold intermediate values as well as register spill (line 1), (b) loading scores through read-only cache instead of shared memory to avoid weak scalability with low occupancy (line 16) and (c) fully unrolling the innermost loop for maximizing registers usage to reside high frequency values on the on-chip memory (line 14). In order to assign different threads of a warp to work on different sequences without mutual interference, we label group ID $gid$ and the offset in group $oig$ on each thread (line 3 and 4). Threads that work on the same sequence are grouped with a unique $gid$, and they are assigned to different in-group tasks based on the $oig$. Inter-thread collaborations are only allowed within each thread group.

\begin{algorithm}
\caption{MSV kernels with multi-sequence alignment}\label{MSVkernel}
\begin{algorithmic}[1]
\Input emission score $E$, sequence data block $B$, height of data block $M$, sequence length $Len$, offset of sequence $O_{seq}$, offset of sequence length $O_{len}$ and other parameters, such as $L$, $W$, $H$, $S$ and $dbias$, etc.
\Output P-values of all sequences $P_i$.
\State local memory $\Gamma[H]$
\State $wid\gets blockIdx.y * blockDim.y + threadIdx.y$
\State $gid\gets \lfloor threadIdx.y/L\rfloor$\Comment{group id}
\State $oig\gets threadIdx.x\ \%\ L$\Comment{offset in group}
\For{$k\gets 0$ \textbf{to} $W-1$}
  \State $R\gets M[wid]$
  \State $count\gets$ \texttt{0}
  \State $mask$, $sc_E\gets$ \texttt{0x00000000}
  \State $I_{seq}\gets O_{seq}[wid]+gid*L+\lfloor k/4\rfloor$
  \State $sc_B\gets$ initialize it based on $Len$, $O_{len}[wid]$ and $k$.
  \While{$count<R$}
    \State $r\gets extract\_res(B,I_{seq},k,S)$
    \State $\gamma\gets inter\_or\_intra\_reorder(L,S,v,oig)$
    \State \#pragma unroll $H$
    \For{$h\gets 0$ \textbf{to} $H-1$}
      \State $\sigma\gets load\_emission\_score(E,S,L,oig,h,r)$
      \State $v\gets \underline{vmaxu4}(\gamma,sc_{B})$
      \State $v\gets \underline{vaddus4}(v,dbias)$
      \State $v\gets \underline{vsubus4}(v,\sigma)$
      \State $sc_{E}\gets \underline{vmaxu4}(sc_{E},v)$
      \State $\gamma\gets\Gamma[h]$\ \texttt{\&} $mask$\Comment{load old scores}
      \State $\Gamma[h]\gets v$\Comment{store new scores}
    \EndFor
    \State $sc_E\gets max\_reduction(L,S,sc_E)$
    \State $sc_J$, $sc_B\gets$ update special states, given $sc_E$.
    \State $mask\gets$ \texttt{0xffffffff}
    \If{$r$ contains ending residue \textbf{@}}\Comment{branch}
      \State $P_i\gets$ calculate P-value and record it.
      \State $mask\gets$ set affected bits to \texttt{0x00}.
      \State $sc_J$, $sc_E$, $sc_B\gets$ update or reset special states.
    \EndIf
    \State $count$, $I_{seq}\gets$ step up.
  \EndWhile
\EndFor
\end{algorithmic}
\end{algorithm}

The outer loop (line 5) iterates over columns of the warp-based sequence data block $B$ while the middle loop (line 11) takes charge of each row. The cycle times of outer loop is directly affected by the query model size: the larger model results in the more cycles. This is because on-chip memory resources are limited when making alignment with large models, and it further leads to the kernel selection with small $S$. For example, a model with length of $45$ can be handled by $128$-lane kernels whereas a model of $1000$-length may only select $4$-lane kernels. Given that, in one warp, $k$ is used to index $S$ columns of residues simultaneously during each iteration, and the $I_{seq}$ always points to residues that are being extracted from global memory. The details of residue extraction are shown in Algorithm~\ref{extractres}. For $S$-lane kernels with $S\leq32$, only one 8-bit value (one residue) per thread group is extracted and be ready to make alignment though a warp always have the fully coalesced access to $128$ residues assembled in $128$-byte memory space. Instead, $64$-lane kernels extract two 8-bit values, and $128$-lane kernels are able to handle all of them. These residues are then used in the function $load\_emission\_score$ (line 16) to load corresponding emission scores of ``Match" states (line 3, 5 and 11 in Algorithm~\ref{loademissionscore}). The total number of amino acids is extended to 32, and the extra states are filled with invalid scores, which aims to cover the newly introduced residues (\textit{ending} and \textit{padding}). $64$ and $128$-lane kernels are treated in a specical way as shown in Algorithm~\ref{loademissionscore} (line 3-9 and line 12-15) due to the demand of score assembly. In this case, each thread assembles two or four scores of different residues into a 32-bit register to be ready for subsequent SIMD instructions. All emission scores are loaded through read-only cache to keep shared/L1 cache path from overuse, and the score sharing is done via inter-thread shuffle instructions.

\begin{algorithm}
\caption{Extract residues from data block\textbf{ - \textit{extract\_res}}}\label{extractres}
\begin{algorithmic}[1]
\Input $B$, $I_{seq}$, $k$ and $S$.
\Output a 32-bit value $r$ that contains one, two or four residues, given $S$.
\If{$S\in\{2,4,8,16,32\}$}
  \State $r\gets (B[I_{seq}]$ \texttt{>>} $8*(k\ \%\ 4))$\ \texttt{\&} \texttt{0x000000ff}
\ElsIf{$S=64$}
  \State $r\gets (B[I_{seq}]$ \texttt{>>} $16*k)$\ \texttt{\&} \texttt{0x0000ffff}
\ElsIf{$S=128$}
  \State $r\gets B[I_{seq}]$
\EndIf
\end{algorithmic}
\end{algorithm}

\begin{algorithm}
\caption{Get ``Match" scores\textbf{ - \textit{load\_emission\_score}}}\label{loademissionscore}
\begin{algorithmic}[1]
\Input $E$, $S$, $L$, $oig$, $h$ and $r$.
\Output a 32-bit value $\sigma$ that contains four emission scores, each is 8-bit, in striped layout.
\State $N_a\gets 32$\Comment{amino acids}
\If{$S\in\{2,4,8,16,32\}$}
  \State $\sigma\gets E[h*N_a*L+r*L+oig]$\Comment{ldg}
\ElsIf{$S=64$}
  \State $sc\gets E[h*N_a+threadIdx.x]$ \texttt{\&} \texttt{0x0000ffff}
  \State $res\gets r$ \texttt{\&} \texttt{0x000000ff}
  \State $\sigma\gets\sigma\ \Vert\ (\_\_shfl(sc,res))$\Comment{assembly}
  \State $res\gets (r$ \texttt{>>} $8)$ \texttt{\&} \texttt{0x000000ff}
  \State $\sigma\gets\sigma\ \Vert\ (\_\_shfl(sc,res)$ \texttt{<<} $16)$\Comment{assembly}
\ElsIf{$S=128$}
  \State $sc\gets E[h*N_a+threadIdx.x]$ \texttt{\&} \texttt{0x000000ff}
  \For{$bits\in\{0,8,16,24\}$}
    \State $res\gets (r$ \texttt{>>} $bits)$ \texttt{\&} \texttt{0x000000ff}
    \State $\sigma\gets\sigma\ \Vert\ (\_\_shfl(sc,res)$ \texttt{<<} $bits)$\Comment{assembly}
  \EndFor
\EndIf
\end{algorithmic}
\end{algorithm}

\begin{figure}[!t]
\includegraphics[width=0.5\textwidth]{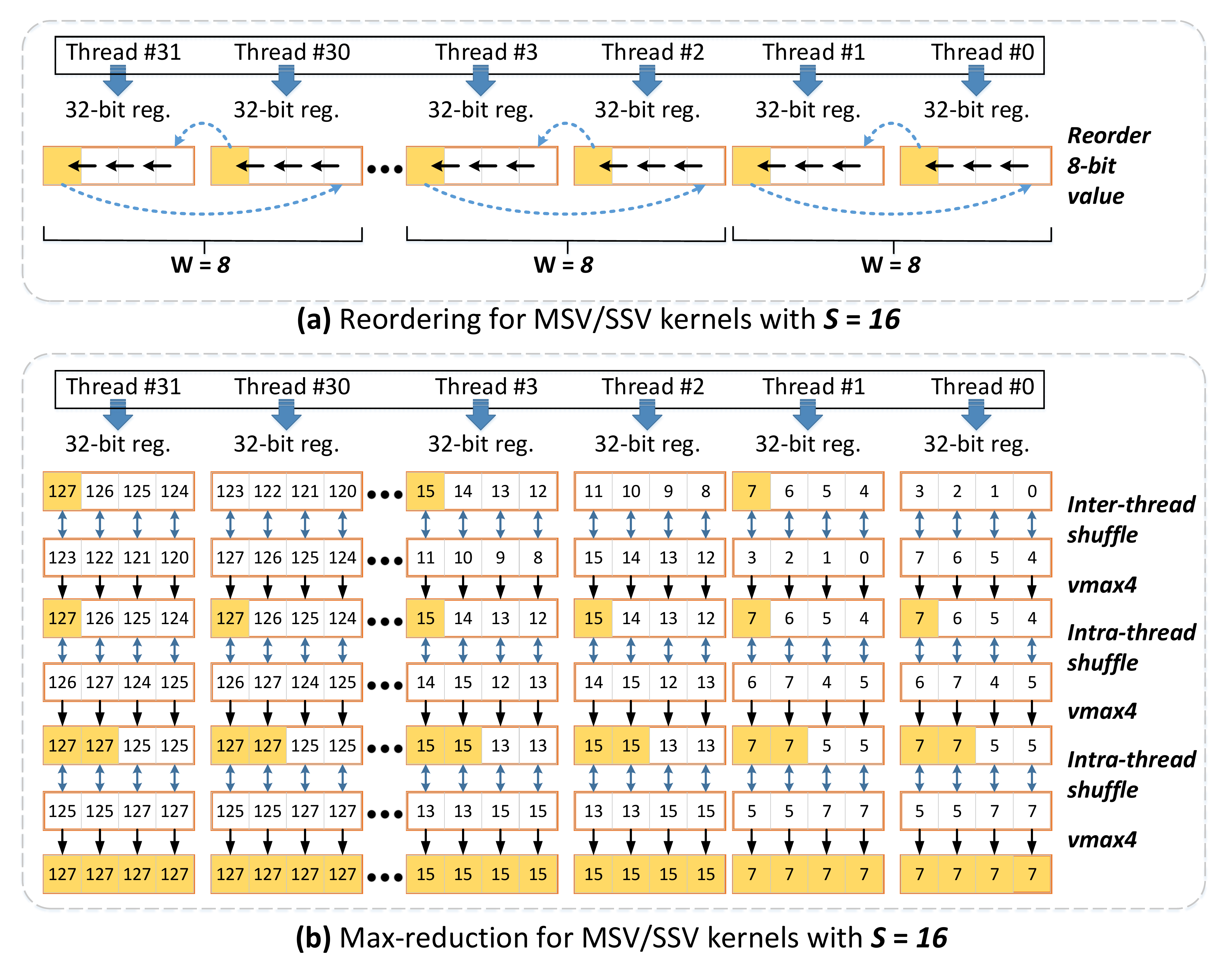}
\centering
\caption{An example of Reordering and Max-reduction for kernels that handle 16 sequences simutaneously.}
\label{16lanekernel}
\end{figure}

Algorithm~\ref{interintrareorder} and~\ref{maxreduction} detail two crucial steps of MSV/SSV kernels, $inter\_or\_intra\_reorder$ and $max\_reduction$, (line 13 and 24 in Algorithm~\ref{MSVkernel}) via the PTX assembly to expose internal mechanisms of massive bitwise operaions for the multi-sequence alignment. They aim to reorder 8-bit values and get the maximum value amongst each thread group in parallel, and meanwhile, noninterference between thread groups is guaranteed. Our design still avoids to use shared memory since available L1 cache is the key factor on performance when unrolling the innermost loop. Therefore, all intermediate or temporary values are held by private memory space of each thread, such as registers or local memory. The shuffle instruction, \texttt{shfl}, is employed again to achieve the inter-thread communication but the difference is that a \texttt{mask} is specified to split a warp into sub-segments (line 8 in Algorithm~\ref{interintrareorder}). Each sub-segment represents a thread group. In Algorithm~\ref{maxreduction}, two reduction phases are required for $S$-lane kernels with $S\leq16$. Line $5$ to $12$ presents inter-thread reductions by using \texttt{shfl} and \texttt{vmax} to get top four values of 8-bit within each thread group. The following lines are intra-thread reductions which only happen inside each thread and eventually works out the maximum value. As an example, Fig.~\ref{16lanekernel} illustrates the reordering and max-reduction for $16$-lane kernels. $W=8$ is the width of striped model states, and it can also be regarded as the scope of each thread group. For $S=16$, a warp is partitioned into $16$ thread groups, and each handles eight values of 8-bit. Yellow lattices in Fig.~\ref{16lanekernel}(a) are values that need to be exchanged betweem two threads. Arrows indicate the reordering direction. In Fig.~\ref{16lanekernel}(b), assuming digital numbers (0 to 127) labeled inside lattices represent the values held by threads, the yellow lattices always track the maximum value of each thread group. Three pairs of shuffle and SIMD instructions are used to calculate the maximum value and broadcast it to all members of the thread group.

\begin{algorithm}
\caption{Reorder 8-bit value inter-thread or intra-thread\textbf{ - \textit{inter\_or\_intra\_reorder}}}\label{interintrareorder}
\begin{algorithmic}[1]
\Input $L$, $S$, $v$ and $oig$.
\Output a 32-bit value $\gamma$ that contain four 8-bit values.
\If{$S\in\{2,4,8,16\}$}\Comment{inter-thread}
  \State $lane\gets (oig + L-1)\ \%\ L$\Comment{source lane}
  \State $asm\{\texttt{shr.u32\ \ }\texttt{x,\ v,\ 24}\}$
  \State $asm\{\texttt{mov.b32\ \ }\texttt{mask,\ 0x1f}\}$
  \State $asm\{\texttt{sub.b32\ \ }\texttt{mask,\ mask,\ L(S)-1}\}$
  \State $asm\{\texttt{shl.b32\ \ }\texttt{mask,\ mask,\ 16}\}$
  \State $asm\{\texttt{or.b32\ \ }\texttt{mask,\ mask,\ 0x1f}\}$
  \State $asm\{\texttt{shfl.idx.b32\ \ }\texttt{x,\ x,\ lane,\ mask}\}$
  \State $asm\{\texttt{shl.b32\ \ }\texttt{y,\ v,\ 8}\}$
  \State $\gamma\gets x\ \vert\ y$
\ElsIf{$S=32$}\Comment{intra-thread}
  \State $asm\{\texttt{shr.u32\ \ }\texttt{x,\ v,\ 24}\}$
  \State $asm\{\texttt{shl.b32\ \ }\texttt{y,\ v,\ 8}\}$
  \State $\gamma\gets x\ \vert\ y$
\ElsIf{$S=64$}\Comment{intra-thread}
  \State $asm\{\texttt{shr.u32\ \ }\texttt{x,\ v,\ 8}\}$
  \State $asm\{\texttt{and.b32\ \ }\texttt{x,\ x,\ 0x00ff00ff}\}$
  \State $asm\{\texttt{shl.b32\ \ }\texttt{y,\ v,\ 8}\}$
  \State $asm\{\texttt{and.b32\ \ }\texttt{y,\ y,\ 0xff00ff00}\}$
  \State $\gamma\gets x\ \vert\ y$
\ElsIf{$S=128$}
  \State $\gamma\gets$ \texttt{0x00000000} or \texttt{0x80808080}\Comment{not reorder}
\EndIf
\end{algorithmic}
\end{algorithm}

\begin{algorithm}
\caption{Get and broadcast maximum value through reduction operations\textbf{ - \textit{max\_reduction}}}\label{maxreduction}
\begin{algorithmic}[1]
\Input $L$, $S$ and $sc_E$.
\Output a 32-bit value $sc_E$ that contains four 8-bit or two 16-bit values.
\If{$S=128$}
  \State do nothing but \textbf{Return} $sc_E$.
\Else
  \State $x\gets sc_E,\ y\gets 0,\ z\gets 0$
  \If{$S\in\{2,4,8,16\}$}\Comment{inter-thread reduction}
    \State $i\gets\log_2{L-1}$
    \For{$lm\gets 2^0$ \textbf{to} $2^i$}
      \State $asm\{\texttt{shfl.bfly.b32\ \ }\texttt{y,\ x,\ lm,\ 0x1f}\}$
      \State $asm\{\texttt{and.b32\ \ }\texttt{m,\ m,\ 0x00000000}\}$
      \State $asm\{\texttt{vmax4.u32.u32.u32\ \ }\texttt{x,\ x,\ y,\ m}\}$
    \EndFor
  \EndIf
  \State $asm\{\texttt{shr.u32\ \ }\texttt{y,\ x,\ 8}\}$
  \State $asm\{\texttt{and.b32\ \ }\texttt{y,\ y,\ 0x00ff00ff}\}$
  \State $asm\{\texttt{shl.b32\ \ }\texttt{z,\ x,\ 8}\}$
  \State $asm\{\texttt{and.b32\ \ }\texttt{z,\ z,\ 0xff00ff00}\}$
  \State $asm\{\texttt{or.b32\ \ }\texttt{y,\ y,\ z}\}$
  \State $asm\{\texttt{and.b32\ \ }\texttt{m,\ m,\ 0x00000000}\}$
  \State $asm\{\texttt{vmax4.u32.u32.u32\ \ }\texttt{x,\ x,\ y,\ m}\}$
  \State \textbf{Return} $sc_E\gets x$ \textbf{if} $S=64$.
  \State $asm\{\texttt{shr.u32\ \ }\texttt{y,\ x,\ 16}\}$
  \State $asm\{\texttt{shl.b32\ \ }\texttt{z,\ x,\ 16}\}$
  \State $asm\{\texttt{or.b32\ \ }\texttt{y,\ y,\ z}\}$
  \State $asm\{\texttt{and.b32\ \ }\texttt{m,\ m,\ 0x00000000}\}$
  \State $asm\{\texttt{vmax4.u32.u32.u32\ \ }\texttt{x,\ x,\ y,\ m}\}$
  \State \textbf{Return} $sc_E\gets x$ \textbf{if} $S\in\{1,2,4,8,16,32\}$.
\EndIf
\end{algorithmic}
\end{algorithm}

The potential divergent execution only happens in the branch for recording P-value of ended sequences, as shown in Algorithm~\ref{MSVkernel}, line 27-31. Threads which find the existence of \textit{ending} residues keep active and move into the branch whereas others are inactive during this period. A \texttt{mask} is introduced to mark the position of ended sequences within 32-bit memory space and set those affected bits to $0$. This is particularly helpful to $64$ and $128$-lane kernels because it only cleans up corresponding lanes for new sequence while keeping data of other lanes unchanged. Moreover, bitwise operations with \texttt{mask} minimize the number of instructions needed inside the innermost loop, which is also beneficial to the overall performance. As for the SSV kernel shown in Algorithm~\ref{SSVkernel}, it shares the same framework with the MSV kernel but has less computational workload. Besides, one more mask value is added to reset affected bits since the \texttt{-inf} of SSV kernel is \texttt{0x80} rather than \texttt{0x00}. $mask1$ cleans up outdated scores as the first step followed by a bitwise disjunction with $mask2$ to reset local memory $\Gamma$ (line 18).

\begin{algorithm}
\caption{SSV kernel with multi-sequence alignment}\label{SSVkernel}
\begin{algorithmic}[1]
\Input emission score $E$, sequence data block $B$, height of data block $M$, sequence length $Len$, offset of sequence $O_{seq}$, offset of sequence length $O_{len}$ and other parameters, such as $L$, $W$, $H$, $S$ and $dbias$, etc.
\Output P-values of all sequences $P_i$.
\State local memory $\Gamma[H]$
\State $wid\gets blockIdx.y * blockDim.y + threadIdx.y$
\State $gid\gets \lfloor threadIdx.y/L\rfloor$\Comment{group id}
\State $oig\gets threadIdx.x\ \%\ L$\Comment{offset in group}
\For{$k\gets 0$ \textbf{to} $W-1$}
  \State $T\gets M[wid]$
  \State $count\gets 0$
  \State $mask1$, $mask2$, $sc_E\gets$ \texttt{0x80808080}
  \State $I_{seq}\gets O_{seq}[wid]+gid*L+\lfloor k/4\rfloor$
  \While{$count<T$}
    \State $r\gets extract\_res(B,I_{seq},k,S)$
    \State $\gamma\gets inter\_or\_intra\_reorder(L,S,v,oig)$
    \State \#pragma unroll $H$
    \For{$h\gets 0$ \textbf{to} $H-1$}
      \State $\sigma\gets load\_emission\_score(E,S,L,oig,h,r)$
      \State $v\gets \underline{vsubus4}(v,\sigma)$
      \State $sc_{E}\gets \underline{vmaxu4}(sc_{E},v)$
      \State $\gamma\gets\Gamma[h]$\ \texttt{\&} $mask1\ \Vert\ mask2$ \Comment{load old scores}
      \State $\Gamma[h]\gets v$\Comment{store new scores}
    \EndFor
    \State $sc_E\gets max\_reduction(L,S,sc_E)$
    \State $mask1\gets$ \texttt{0xffffffff}, $mask2\gets$ \texttt{0x00000000}
    \If{$r$ contains ending residue \textbf{@}}\Comment{branch}
      \State $P_i\gets$ calculate P-value and record it.
      \State $mask1\gets$ set affected bits to \texttt{0x00}.
      \State $mask2\gets$ set affected bits to \texttt{0x80}.
      \State $sc_E\gets$ update or reset special states.
    \EndIf
    \State $count, I_{seq}\gets$ step up.
  \EndWhile
\EndFor
\end{algorithmic}
\end{algorithm}

\subsection{Kernel Optimization}
\label{kernelopt}
The kernel performance of CUDAMPF shows that $H=19$ is able to cover the longest query model whose length is $2405$, for MSV and SSV algorithms~\cite{Jiang2016CUDAMPF}. In current design, however, $2$-lane kernels can only handle models with the length of $1216$ at most, given the same $H$. In addition, we recall that L1 Cache-Hit-Ratio (CHR) is employed as a metric to evaluate register spill in CUDAMPF, and MSV/SSV kernels with maximum 64 registers per thread have no spill to local memory. This enables us to push up $H$ to hold larger models for the multi-sequence alignment. Two optimization schemes are proposed to improve overall performance and address the concern about scalability.

\subsubsection{CHR Sacrificed Kernel}
\label{CHRopt}
It is well-known that L1 cache shares the same block of on-chip memory with shared memory physically, and it is used to cache accesses to local memory as well as register spill. The L1 cache has low latency on data retrieval and storage with a cache hit, which can be further utilized to increase the throughput of kernels based on the proposed framework. We treat L1 cache as \textit{secondary registers}, and the usage is measured by CHR for local loads and stores. By increasing up $H$, more model states can reside in registers and cached local memory. The moderate loss of performance due to uncached register spills are acceptable, which is attributed to highly optimized task and data parallelism in the current framework. However, it is impossible to increase $H$ unboundedly due to the limited capacity of L1 cache. Overly large $H$ leads to the severe register spill that causes low CHR and stalls warps significantly. Hence, there always is a trade-off between CHR and $H$, and the goal is to find a reasonable point where kernel performance (GCUPS) begins fall off. The decline in performance indicates that the latency, caused by excessive communications between on and off-chip memory, starts to overwhelm the benefits of parallelism. The corresponding $H$ at the turning point is considered to be the maximum one, denoted by $H_{max}$.

\begin{table}[h!]
  \caption{Benchmarks of the maximum $H$ via innermost loop unrolling}              
  \centering
  \begin{threeparttable}                                  
      \begin{tabular}{c c c c c c c}                        
        \toprule
        \thead{$H$}&\thead{reg.\\per thread}&\thead{stack\\frame}&\thead{spill\\stores}&\thead{spill\\loads}&\thead{GCUPS}&\thead{L1 CHR\\(\%)}\\
        \midrule
        \multicolumn{7}{c}{\textbf{MSV kernels}}\\
        \midrule
        20 & 63 & 8 & 0 & 0 & 258.7 & 99.97\\  
        25 & 63 & 8 & 0 & 0 & 261.5 & 99.97\\
        30 & 64 & 8 & 0 & 0 & 272.1 & 99.97\\
        35 & 64 & 40 & 40 & 44 & 279.3 & 75.65\\
        40 & 64 & 48 & 52 & 56 & 283.9 & 75.41\\
        \rowcolor{Gray}
        \textbf{45} & \textbf{64} & \textbf{48} & \textbf{56} & \textbf{52} & \textbf{280.6} & \textbf{75.49}\\
        50 & 64 & 96 & 152 & 96 & 263.9 & 25.61\\
        55 & 64 & 128 & 208 & 140 & 240.8 & 14.95\\
        \midrule
        \multicolumn{7}{c}{\textbf{SSV kernels}}\\
        \midrule
        20 & 62 & 8 & 0 & 0 & 269.5 & 99.96\\
        25 & 62 & 8 & 0 & 0 & 314.0 & 99.96\\
        30 & 62 & 8 & 0 & 0 & 329.0 & 99.96\\
        35 & 61 & 8 & 0 & 0 & 347.9 & 99.96\\
        40 & 64 & 16 & 4 & 4 & 361.5 & 99.94\\
        45 & 64 & 48 & 44 & 40 & 375.9 & 80.44\\
        \rowcolor{Gray}
        \textbf{50} & \textbf{64} & \textbf{56} & \textbf{64} & \textbf{44} & \textbf{375.9} & \textbf{60.30}\\
        55 & 64 & 80 & 116 & 72 & 340.9 & 17.81\\
        \bottomrule
      \end{tabular}
      \begin{tablenotes}[flushleft]
        \scriptsize
        \item[***] Data collections on 32-lane kernels compiled with \textit{nvcc} 8.0. Use \textit{env\_nr}~\cite{protein2014dataset} as the sequence dataset.
      \end{tablenotes}
  \end{threeparttable}\label{unrollmax}
\end{table}

TABLE~\ref{unrollmax} lists a benchmark result that shows the relationship between $H$, kernel performance and CHR. Starting from $H=20$ with a step of $5$ , intuitively, CHR is being consumed after on-chip registers are exhausted, and the kernel performance increases first and falls back eventually as expected. We choose $45$ and $50$ as the $H_{max}$ for MSV and SSV kernels, respectively. Larger $H$ results in rapid degradation of both performance and L1 CHR. Besides, the difference of $H_{max}$ indicates that MSV kernels have more instructions than SSV kernels within the innermost loop, and hence more registers or local memory are used while unrolling the loop. The $H_{max}$ is therefore algorithm-dependent. Given Eq.~(\ref{svaluerange}),~(\ref{othervalue}) and $H_{max}$, we formulate the selection of $S$ as below:
\begin{equation}\label{sselection}
\argmax_S \boldsymbol{f}=\{S\mid \boldsymbol{f}=W_SH_{max},\forall\hat{m}\leq W_2H_{max}\colon\boldsymbol{f}\geq\hat{m}\},
\end{equation}
where $\boldsymbol{f}$ is the function of $S$, and $W_2H_{max}$ indicates the maximum length of query models that $2$-lane kernels can handle. The CUDAMPF implementation will be used instead if any larger model is applicable. Eq.~(\ref{sselection}) describes a rule of kernel selection that always prefer to use kernels with more lanes if they are able to cover the model length. Once the kernel type ($S$-lane) is determined, the $H$ of every query model located in the coverage area can be obtained via:
\begin{multline}\label{hselection}
\argmin_H\boldsymbol{\Phi}=\{H\mid\forall H\in[\lceil\frac{\hat{m}}{W_S}\rceil,H_{max}]\cap\mathbb{Z}\colon\boldsymbol{\Phi}=W_SH,\\
\boldsymbol{\Phi}\geq\hat{m}\},
\end{multline}
where $\boldsymbol{\Phi}$ represents the function of $H$. Eq.~(\ref{hselection}) minimizes $H$ to fit query models perfectly, which thereby avoid redundant computation and memory allocation.

In summary, this optimization scheme aims to fully leverage the speedy on-chip memory, including L1 cache via sacrificing CHR proactively, to further boost kernel throughput, and in the meanwhile, it extends coverage of the proposed framework to larger query models.


\subsubsection{Performance-oriented Kernel}
\label{switchopt}
Although the proposed framework achieves a significant improvement in performance, it certainly introduces overhead due to the implementation of multi-sequence alignment, compared with CUDAMPF. This downside becomes more apparent as the model length increases ($S$ decreases). Therefore, it is expected that CUDAMPF with single-sequence alignment may exceed CUDAMPF++ for large enough query models. In order to pursue optimal performance consistently, we also merge CUDAMPF implemention into the proposed framework as a special case with $S\;$=$\;1$. The maximum model length $\hat{m}_{max}$ on which CUDAMPF++ still outperforms is defined as the threshold of kernel switch.

Similar to $\hat{m}_{max}$, another threshold $\hat{m}_{min}$ can also be employed to optimize 128 and 64-lane kernels for small models. We recall that 128 and 64-lane kernels need extra operations to load emission scores in Algorithm~\ref{loademissionscore}. Thus, they have more overhead than other kernels within the innermost loop, which may counteract their advantages on the number of parallel lanes. We extend the coverage of 32-lane kernels to handle small models owned by 128 and 64-lane kernels previously, and the evaluation is presented in Sec.~\ref{scalabilityeval}.

\section{Experimental Results}
In this section, we present several performance evaluations on the proposed CUDAMPF++, such as workload balancing, kernel throughput and scalability. The comparison targets consist of CUDAMPF, CPU-based \textit{hmmsearch} of latest HMMER v3.1b2 and other related acceleration attempts. Both CUDAMPF++ and CUDAMPF are evaluated on a NVIDIA Tesla K40 GPU and compiled with CUDA v8.0 compiler. Tesla K40 is built with the Kepler GK110 architecture that contains 15 SMXs (2880 CUDA cores) and 12 GB off-chip memory~\cite{teslafamily}. One of NVIDIA profiling tools, \textit{nvprof}~\cite{nvprof}, is also used to track metrics like L1/tex CHR, register usage and spill. For \textit{hmmsearch}, two types of CPUs are employed to collect performance results: Intel Xeon E5620 (4 physical cores with maximum 8 threads) and dual Intel Xeon E5-2650 (16 physical cores and 32 threads in total). All programs are executed in the 64-bit Linux operating system.

Unlike CUDAMPF implementation, the NVRTC is deprecated in current design due to its unstability in compiling kernels with high usage of on-chip registers. Even with latest compiler \textit{nvcc} v8.0, runtime compilation with the NVRTC library still generates unexpected binary files or report the error of resources exhaustion, especially when unrolling large loops and register spill happens. Thus, we choose the just-in-time (JIT) compilation instead. All kernels are pre-compiled in offline mode and stored as \textit{.ptx} files, each with an unique kernel name inside. Given different query models, the corresponding \textit{.ptx} file is loaded and further compiled to binary code at runtime. The load time and overhead of compilation are negligible.

Two protein sequence databsets~\cite{protein2014dataset} are chosen for experiments: (a) \textit{env\_nr} (1.9 GB) and (b) \textit{est\_human} (5.6 GB). As for query models, we still use Pfam 27.0~\cite{pfam27} that contains 34 thousand HMMs with different sizes ranging from 7 to 2405. The overall performance is measured in kernel throughput (GCUPS) which is directly calculated by the total number of residues contained in each database, model length and kernel execution time.

\subsection{Evaluation of Workload Balancing}
\label{workloadeval}
To avoid time overhead of data reformatting introduced in Sec.~\ref{dataformat}, we incorporate Redis~\cite{redisbook}, a high performance in-memory database, into the proposed framework. Redis is written in ANSI C and able to work with CUDA seamlessly. It currently works as an auxiliary component to hold warp-based sequence data blocks and query models in memory, which offers blazing fast speed for data retrieval. Given the single K40 GPU, each protein sequence dataset is partitioned into 61,440 blocks which are then ingested into Redis database separately as key-value pairs. The quantity of data blocks resided in Redis database should be integral multiple of the number of available warps.


\begin{table*}[h!]
  \caption{Evaluation of workload balancing for the warp-based sequence data blocks}       
  \centering
  \begin{threeparttable}                                  
      \begin{tabular}{c c c c c c c c c c c}              
        \toprule
        \thead{DB\\name}&\thead{DB\\size (GB)}&\thead{total\\seq.}&\thead{total\\residues}&\thead{avg. $M$}&\thead{sd. $M$}&\thead{avg. \textit{ending}\\residues}&\thead{sd. \textit{ending}\\residues}&\thead{avg. \textit{PRR}}&\thead{SMX eff.\\MSV (\%)\tnote{*}}&\thead{SMX eff.\\SSV (\%)\tnote{*}}\\
        \midrule
        env\_nr & 1.9 & 6,549,721 & 1,290,247,663 & 21,109 & 85 & 13,645 & 71 & 1.14E-4 & 97.13 & 96.7\\
        \midrule
        est\_human & 5.6 & 8,704,954 & 4,449,477,543 & 72,563 & 174 & 18,135 & 60 & 3.22E-5 & 97.97 & 97.76\\
        \bottomrule
      \end{tabular}
      \begin{tablenotes}[flushleft]
        \scriptsize
        \item[*] Data collection with 32-lane kernels.
      \end{tablenotes}
  \end{threeparttable}\label{tb:workloadeval}
\end{table*}

Table~\ref{tb:workloadeval} summarizes the evaluation result of workload balancing for both protein sequence datasets. The ``avg." and ``sd." represent average value and standard deviation across all blocks, respectively. We recall that $M$ is the height of data block which serves as the metrics of computational workload for each warp, and the number of \textit{ending} residues is another impact factor of performance because the \textit{ending} residue may lead to thread idling. It is clear to see that both sd.~$M$ and sd.~\textit{ending} residues are trivial, and the last two columns show that average multiprocessor efficiency approaches 100\%, which are strong evidences of balanced workload over all warps on GPU. Besides, the Padding-to-Real Ratio (\textit{PRR}) that compares the level of invalid computation to the level of desired computation is investigated to assess the negative effect of padding residues, and it is also proved to be negligible.

\subsection{Performance Evaluation and Analysis}
\label{performanceeval}
In order to demonstrate the outstanding performance of proposed method and its correlation with the utilization of memory resources, we make an in-depth comparison between CUDAMPF++ and CUDAMPF via profiling both MSV and SSV kernels, reported in Table~\ref{tb:msvcomparison} and~\ref{tb:ssvcomparison}, respectively. A total of $27$ query models are selected to investigate the impact of $H$ on the performance of $S$-lane kernels. Each kernel type is evaluated with two models that correspond to $H\;$=$\;H_{max}$ and $H\;$=$\;\lceil\frac{H_{max}-1}{2}+1\rceil$, except for the $2$-lane kernel since the $2405$ is the largest model length in~\cite{pfam27} with corresponding $H\;$=$\;38$.

\begin{table*}[h!]
  \caption{Performance comparison of MSV kernel between the proposed CUDAMPF++ and CUDAMPF}       
  \centering
  \begin{threeparttable}                                  
      \begin{tabular}{c c c c c c c c c c c c}              
        \toprule
        & & \multicolumn{9}{c}{CUDAMPF++ \textbf{\textit{vs.}} CUDAMPF}\\
        \cmidrule{4-12}
        \thead{$S$-lane\\kernels}&\thead{model\\length $\hat{m}$}&\thead{acc. ID}&\thead{$H$}&\thead{reg.\\per thread}&\thead{stack\\frame}&\thead{spill\\stores}&\thead{spill\\loads}&\thead{L1 CHR\\(\%)}&\thead{Tex. CHR\\(\%)}&\thead{GCUPS}&\thead{speedup}\\
        \midrule
        128 & 23 & PF13823.1 & 23 \textbf{/} 2 & 63 \textbf{/} 29 & 24 \textbf{/} 0 & 0 \textbf{/} 0 & 0 \textbf{/} 0 & 99.94 \textbf{/} \textit{unused} & 100 \textbf{/} 100 & 168.6 \textbf{/} 9.9 & 17.0x\\
        128 & 45 & PF05931.6 & 45 \textbf{/} 2 & 64 \textbf{/} 29 & 48 \textbf{/} 0 & 48 \textbf{/} 0 & 24 \textbf{/} 0 & 51.93 \textbf{/} \textit{unused} & 100 \textbf{/} 100 & 144.4 \textbf{/} 19.4 & 7.5x\\
        64 & 46 & PF09501.5 & 23 \textbf{/} 2 & 64 \textbf{/} 29 & 16 \textbf{/} 0 & 0 \textbf{/} 0 & 0 \textbf{/} 0 & 99.96 \textbf{/} \textit{unused} & 100 \textbf{/} 100 & 225.7 \textbf{/} 19.8 & 11.4x\\
        64 & 90 & PF05777.7 & 45 \textbf{/} 2 & 64 \textbf{/} 29 & 48 \textbf{/} 0 & 64 \textbf{/} 0 & 32 \textbf{/} 0 & 50.88 \textbf{/} \textit{unused} & 100 \textbf{/} 100 & 231.3 \textbf{/} 38.7 & 6.0x\\
        32 & 92 & PF00207.17 & 23 \textbf{/} 2 & 64 \textbf{/} 29 & 8 \textbf{/} 0 & 0 \textbf{/} 0 & 0 \textbf{/} 0 & 99.97 \textbf{/} \textit{unused} & 100 \textbf{/} 100 & 274.1 \textbf{/} 39.6 & 7.0x\\
        32 & 180 & PF02737.13 & 45 \textbf{/} 2 & 64 \textbf{/} 29 & 48 \textbf{/} 0 & 56 \textbf{/} 0 & 52 \textbf{/} 0 & 75.49 \textbf{/} \textit{unused} & 100 \textbf{/} 100 & 278.8 \textbf{/} 77.4 & 3.6x\\
        16 & 184 & PF00596.16 & 23 \textbf{/} 2 & 63 \textbf{/} 29 & 8 \textbf{/} 0 & 0 \textbf{/} 0 & 0 \textbf{/} 0 & 99.99 \textbf{/} \textit{unused} & 100 \textbf{/} 100 & 266.7 \textbf{/} 78.9 & 3.4x\\
        16 & 360 & PF01117.15 & 45 \textbf{/} 3 & 64 \textbf{/} 30 & 56 \textbf{/} 0 & 60 \textbf{/} 0 & 60 \textbf{/} 0 & 80.11 \textbf{/} \textit{unused} & 100 \textbf{/} 100 & 277.0 \textbf{/} 130.9 & 2.1x\\
        8 & 368 & PF05208.8 & 23 \textbf{/} 3 & 63 \textbf{/} 30 & 8 \textbf{/} 0 & 0 \textbf{/} 0 & 0 \textbf{/} 0 & 99.99 \textbf{/} \textit{unused} & 100 \textbf{/} 100 & 266.7 \textbf{/} 133.6 & 2.0x\\
        8 & 720 & PB000053 & 45 \textbf{/} 6 & 64 \textbf{/} 34 & 56 \textbf{/} 0 & 60 \textbf{/} 0 & 60 \textbf{/} 0 & 80.10 \textbf{/} \textit{unused} & 78.48 \textbf{/} 92.37 & 271.6 \textbf{/} 183.8 & 1.5x\\
        4 & 735 & PF03971.9 & 23 \textbf{/} 6 & 63 \textbf{/} 34 & 8 \textbf{/} 0 & 0 \textbf{/} 0 & 0 \textbf{/} 0 & 100 \textbf{/} \textit{unused} & 75.3 \textbf{/} 92.37 & 262.4 \textbf{/} 187.4 & 1.4x\\
        4 & 1439 & PF12252.3 & 45 \textbf{/} 12 & 64 \textbf{/} 51 & 56 \textbf{/} 0 & 60 \textbf{/} 0 & 60 \textbf{/} 0 & 80.08 \textbf{/} \textit{unused} & 67.09 \textbf{/} 72.66 & 271.5 \textbf{/} 231.6 & 1.2x\\
        2 & 1471 & PB006678 & 23 \textbf{/} 12 & 63 \textbf{/} 51 & 8 \textbf{/} 0 & 0 \textbf{/} 0 & 0 \textbf{/} 0 & 100 \textbf{/} \textit{unused} & 63.38 \textbf{/} 72.6 & 259.7 \textbf{/} 237.3 & 1.1x\\
        2 & 2405 & PB003055 & 38 \textbf{/} 19 & 64 \textbf{/} 64 & 40 \textbf{/} 0 & 32 \textbf{/} 0 & 44 \textbf{/} 0 & 100 \textbf{/} \textit{unused} & 60.15 \textbf{/} 64.06 & 264.0 \textbf{/} 271.2 & -1.0x\\
        \bottomrule
      \end{tabular}
      \begin{tablenotes}[flushleft]
        \scriptsize
        \item[***] The \textit{env\_nr}~\cite{protein2014dataset} is used in data collection.
      \end{tablenotes}
  \end{threeparttable}\label{tb:msvcomparison}
\end{table*}

For the MSV kernels, the maximum speedup listed on Table~\ref{tb:msvcomparison} is $17.0$x when $\hat{m}\;$=$\;23$, and the trend of speedup is descending as model length increases. This is because memory resources, like on-chip registers, L1 cache and even local memory, are significantly underutilized in CUDAMPF when making alignment with small models whereas CUDAMPF++ always intends to fully take advantage of them. Given $64$ as the maximum number of register per thread, only about half the amount of registers are occupied in CUDAMPF till $\hat{m}\;$=$\;735$, and other resources are not utilized at all. In contrast, the CUDAMPF++ not only keeps high usage of registers but also utilizes L1 cache and local memory to hold more data, which results in a near constant performance regardless of the model length. The texture CHR is dominated by model length since we only use texture cache for loading emission scores. Larger model leads to lower texture CHR. Comparing the performance of $S$-lane kernels in CUDAMPF++, the cases of $H\;$=$\;45$ outperform the cases of $H\;$=$\;23$ though more local memory are allocated with register spill. One exception is the $128$-lane kernel due to its higher complexity of innermost loop, which can be optimized via using the $32$-lane kernel instead.

\begin{table*}[h!]
  \caption{Performance comparison of SSV kernel between the proposed CUDAMPF++ and CUDAMPF}       
  \centering
  \begin{threeparttable}                                  
      \begin{tabular}{c c c c c c c c c c c c}              
        \toprule
        & & \multicolumn{9}{c}{CUDAMPF++ \textbf{\textit{vs.}} CUDAMPF}\\
        \cmidrule{4-12}
        \thead{$S$-lane\\kernels}&\thead{model\\length}&\thead{acc. ID}&\thead{$H$}&\thead{reg.\\per thread}&\thead{stack\\frame}&\thead{spill\\stores}&\thead{spill\\loads}&\thead{L1 CHR\\(\%)}&\thead{Tex. CHR\\(\%)}&\thead{GCUPS}&\thead{speedup}\\
        \midrule
        128 & 26 & PF02822.9 & 26 \textbf{/} 2 & 62 \textbf{/} 33 & 24 \textbf{/} 0 & 0 \textbf{/} 0 & 0 \textbf{/} 0 & 99.93 \textbf{/} \textit{unused} & 100 \textbf{/} 100 & 178.4 \textbf{/} 15.9 & 11.2x\\
        128 & 50 & PF03869.9 & 50 \textbf{/} 2 & 64 \textbf{/} 33 & 56 \textbf{/} 0 & 60 \textbf{/} 0 & 32 \textbf{/} 0 & 54.29 \textbf{/} \textit{unused} & 100 \textbf{/} 100 & 144.6 \textbf{/} 30.7 & 4.7x\\
        64 & 52 & PF02770.14 & 26 \textbf{/} 2 & 63 \textbf{/} 33 & 16 \textbf{/} 0 & 0 \textbf{/} 0 & 0 \textbf{/} 0 & 99.96 \textbf{/} \textit{unused} & 100 \textbf{/} 100 & 276.1 \textbf{/} 31.8 & 8.7x\\
        64 & 100 & PB000229 & 50 \textbf{/} 2 & 64 \textbf{/} 33 & 72 \textbf{/} 0 & 100 \textbf{/} 0 & 52 \textbf{/} 0 & 30.65 \textbf{/} \textit{unused} & 100 \textbf{/} 100 & 239.8 \textbf{/} 61.4 & 3.9x\\
        32 & 104 & PF14807.1 & 26 \textbf{/} 2 & 61 \textbf{/} 33 & 8 \textbf{/} 0 & 0 \textbf{/} 0 & 0 \textbf{/} 0 & 99.96 \textbf{/} \textit{unused} & 100 \textbf{/} 100 & 307.8 \textbf{/} 63.6 & 4.8x\\
        32 & 200 & PF13087.1 & 50 \textbf{/} 2 & 64 \textbf{/} 33 & 56 \textbf{/} 0 & 64 \textbf{/} 0 & 44 \textbf{/} 0 & 60.3 \textbf{/} \textit{unused} & 100 \textbf{/} 100 & 365.0 \textbf{/} 122.4 & 3.0x\\
        16 & 208 & PF15420.1 & 26 \textbf{/} 2 & 62 \textbf{/} 33 & 8 \textbf{/} 0 & 0 \textbf{/} 0 & 0 \textbf{/} 0 & 99.98 \textbf{/} \textit{unused} & 100 \textbf{/} 100 & 305.0 \textbf{/} 127.2 & 2.4x\\
        16 & 400 & PF13372.1 & 50 \textbf{/} 4 & 64 \textbf{/} 37 & 56 \textbf{/} 0 & 72 \textbf{/} 0 & 52 \textbf{/} 0 & 58.57 \textbf{/} \textit{unused} & 95.66 \textbf{/} 100 & 347.8 \textbf{/} 199.3 & 1.7x\\
        8 & 416 & PF06808.7 & 26 \textbf{/} 4 & 62 \textbf{/} 37 & 8 \textbf{/} 0 & 0 \textbf{/} 0 & 0 \textbf{/} 0 & 99.99 \textbf{/} \textit{unused} & 98.92 \textbf{/} 100 & 302.6 \textbf{/} 209.1 & 1.4x\\
        8 & 800 & PF02460.13 & 50 \textbf{/} 7 & 64 \textbf{/} 42 & 56 \textbf{/} 0 & 72 \textbf{/} 0 & 52 \textbf{/} 0 & 58.53 \textbf{/} \textit{unused} & 75.79 \textbf{/} 87.24 & 344.8 \textbf{/} 306.9 & 1.1x\\
        4 & 832 & PB001474 & 26 \textbf{/} 7 & 62 \textbf{/} 42 & 8 \textbf{/} 0 & 0 \textbf{/} 0 & 0 \textbf{/} 0 & 99.99 \textbf{/} \textit{unused} & 76.90 \textbf{/} 87.24 & 302.1 \textbf{/} 319.1 & -1.1x\\
        4 & 1600 & PB000744 & 50 \textbf{/} 13 & 64 \textbf{/} 56 & 56 \textbf{/} 0 & 72 \textbf{/} 0 & 52 \textbf{/} 0 & 58.49 \textbf{/} \textit{unused} & 65.68 \textbf{/} 70.80 & 349.3 \textbf{/} 403.5 & -1.2x\\
        2 & 1630 & PB000663 & 26 \textbf{/} 13 & 62 \textbf{/} 56 & 8 \textbf{/} 0 & 0 \textbf{/} 0 & 0 \textbf{/} 0 & 100 \textbf{/} \textit{unused} & 66.57 \textbf{/} 70.81 & 293.1 \textbf{/} 411.1 & -1.4x\\
        2 & 2405 & PB003055 & 38 \textbf{/} 19 & 64 \textbf{/} 63 & 16 \textbf{/} 0 & 0 \textbf{/} 0 & 0 \textbf{/} 0 & 100 \textbf{/} \textit{unused} & 59.69 \textbf{/} 64.05 & 318.8 \textbf{/} 468.9 & -1.5x\\
        \bottomrule
      \end{tabular}
      \begin{tablenotes}[flushleft]
        \scriptsize
        \item[***] The \textit{env\_nr}~\cite{protein2014dataset} is used in data collection.
      \end{tablenotes}
  \end{threeparttable}\label{tb:ssvcomparison}
\end{table*}

As shown in Table~\ref{tb:ssvcomparison}, SSV kernels have similar evaluation results with MSV kernels but higher throughput. Starting from $\hat{m}\;$=$\;832$, nevertheless, CUDAMPF outperforms and eventually yields upto 468.9 GCUPS which is $1.5$x faster than CUDAMPF++. The case that peak performance of two frameworks are not comparable is due to the overhead of extra instructions introduced for the multi-sequence alignment in CUDAMPF++. The kernel profiling indicates that both MSV and SSV kernels are bounded by computation and memory bandwidth (texture). However, unlike MSV kernels, SSV kernels have fewer operations within the innermost loop, which makes them more ``sensitive". In other words, newly added operations (i.e., bitwise operations for mask) within the innermost loop, compared with CUDAMPF, have more negative effect on SSV kernels than MSV kernels. Therefore, an upto 50\% performance gap is observed only in SSV kernels.

\subsection{Scalability Evaluation}
\label{scalabilityeval}
\begin{figure*}[!h]
\includegraphics[width=\textwidth]{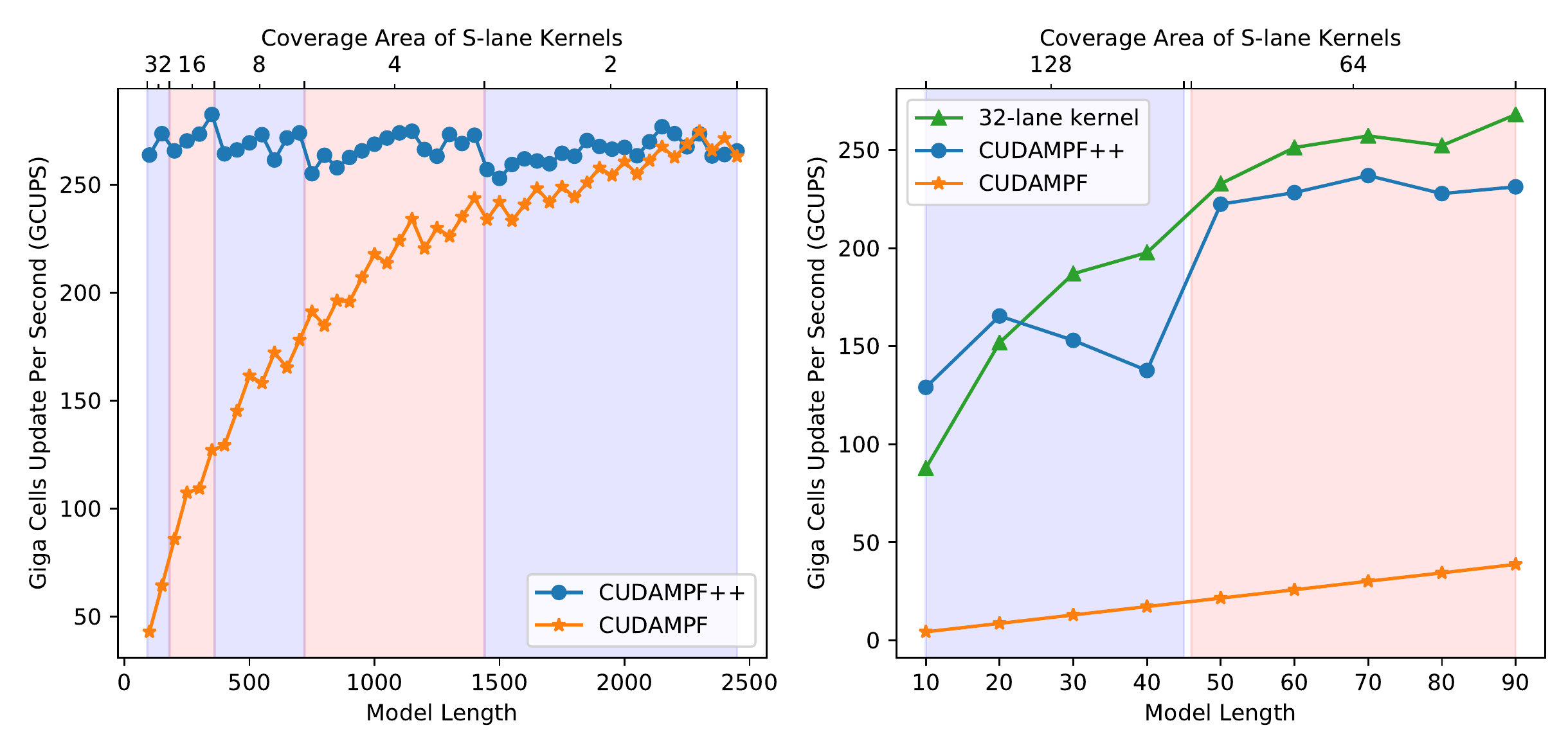}
\centering
\caption{Performance comparison between CUDAMPF++ and CUDAMPF for the MSV kernel.}
\label{fig:msvallcompare}
\end{figure*}

\begin{figure*}[!h]
\includegraphics[width=\textwidth]{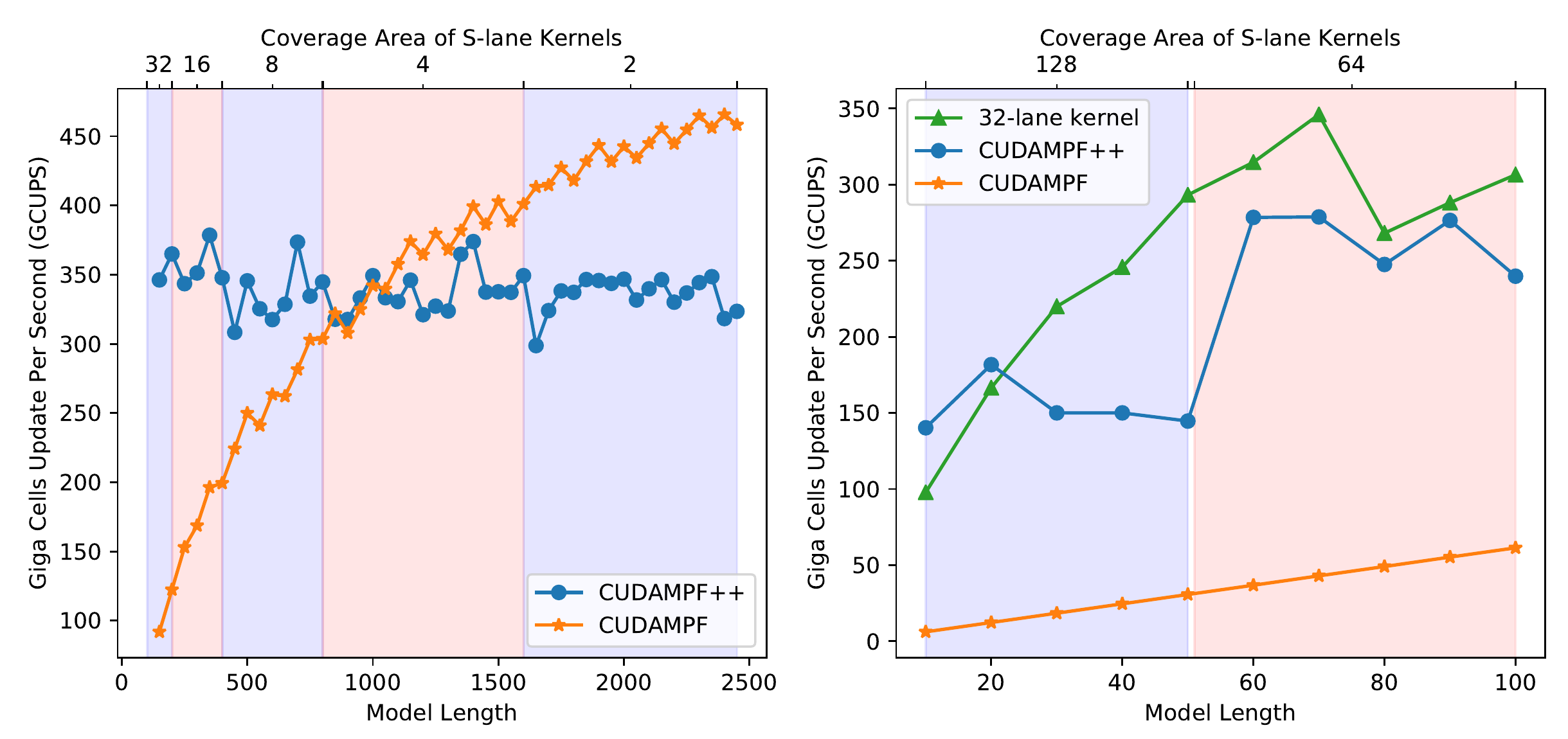}
\centering
\caption{Performance comparison between CUDAMPF++ and CUDAMPF for the SSV kernel.}
\label{fig:ssvallcompare}
\end{figure*}

In order to demonstrate the scalability of the proposed framework, a total of $57$ query models with different sizes ranging from $10$ to $2450$ are investigated. The interval of model length is fixed to $50$. Fig.~\ref{fig:msvallcompare} and~\ref{fig:ssvallcompare} show the performance comparison between CUDAMPF++ and CUDAMPF for MSV and SSV kernels, respectively. The coverage area of model length for each kernel type is highlighted. Right subfigure depicts the performance of 128 and 64-lane kernels while others are shown in the left one. Overall, CUDAMPF++ achieves near constant performance and significantly outperform CUDAMPF with small models. The $S$-lane MSV (SSV) kernel yields the maximum speedup of $30$x ($23$x) with respect to CUDAMPF. It is worth mentioning that, in CUDAMPF++, SSV kernels have larger fluctuation margin of performance than MSV kernels. This is caused by the overhead of using read-only cache (texture cache) to load emission scores. Although both MSV and SSV kernels have only one \textit{\_\_ldg()} function inside innermost loop, the texture cache read in SSV kernels, as one of kernel limits, has a higher proportion of negative effect on performance than that in MSV kernels, which results in such obvious fluctuation. A simple evidence is that the performance curve will be smooth and regular if replacing texture cache reads with a constant value.


Besides, 32-lane kernels are also tested to compare with 128 and 64-lane kernels. By decreasing $H$, the 32-lane kernel is able to cover smaller query models, and it outperforms 128 and 64-lane kernels until the model length is slightly larger than $20$. We simply set $\hat{m}_{min}\;$=$\;20$ for both MSV and SSV kernels in terms of evaluation results. As for $\hat{m}_{max}\;$, the model lengths of $2450$ and $1000$ are selected for MSV and SSV kernels, respectively.

\subsection{Performance Comparison: CUDAMPF++ vs. Others}
\label{overallcompare}
\begin{figure*}[!h]
\includegraphics[width=\textwidth]{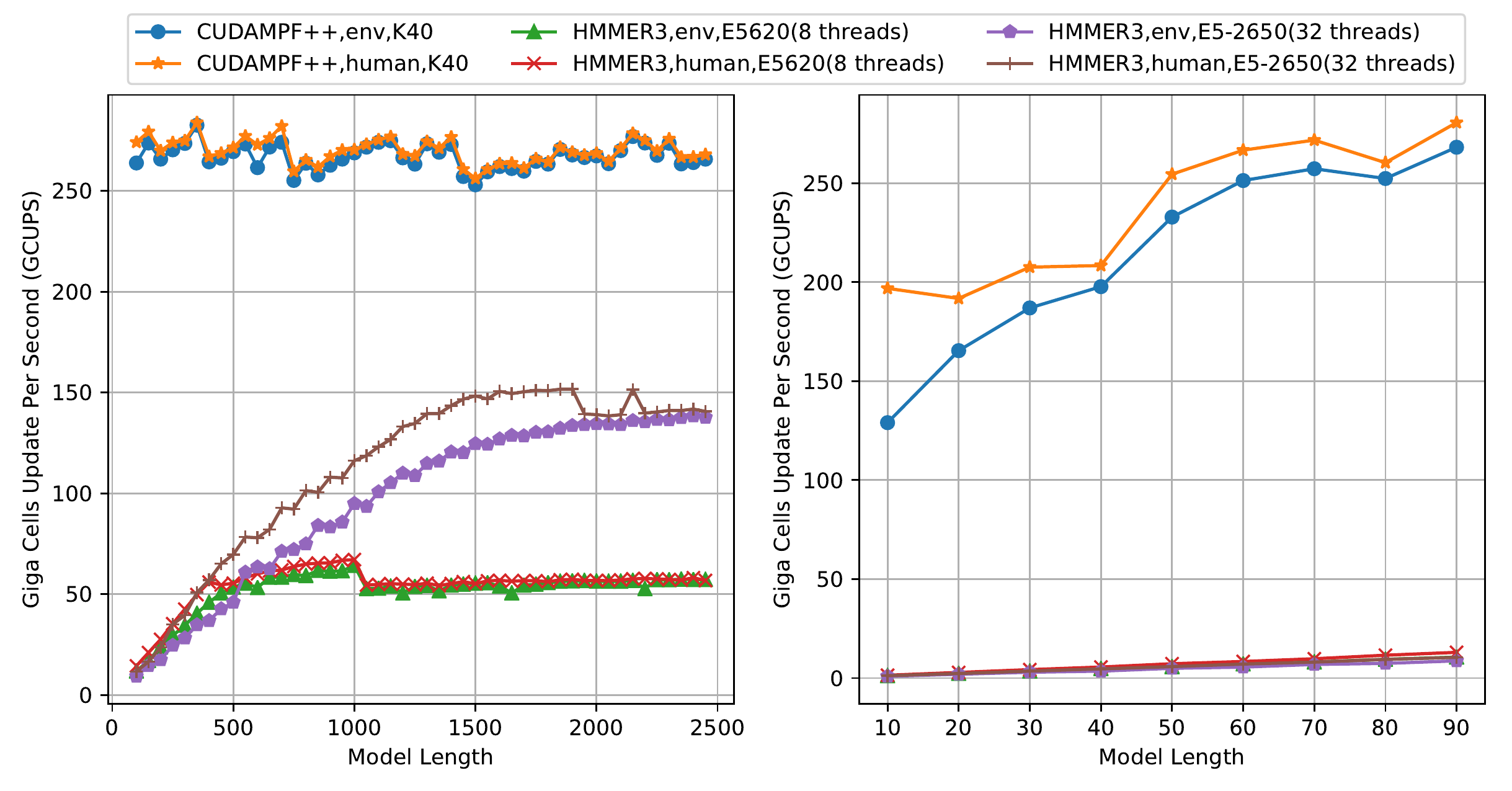}
\centering
\caption{Performance comparison between CUDAMPF++ and HMMER3's CPU-based implementation for the MSV kernel (stage).}
\label{fig:msvcomparehmmer}
\end{figure*}

\begin{figure*}[!h]
\includegraphics[width=\textwidth]{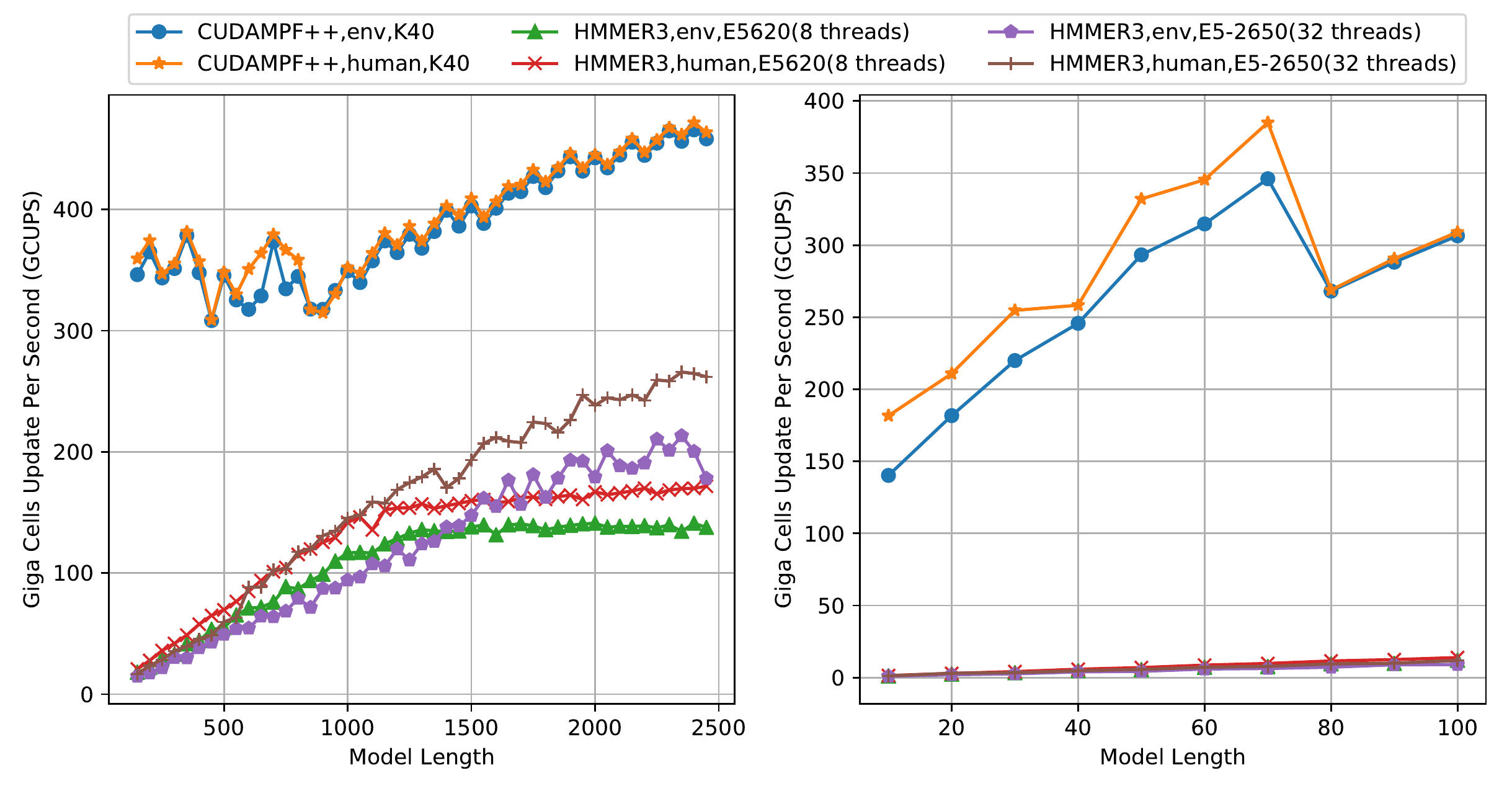}
\centering
\caption{Performance comparison between CUDAMPF++ and HMMER3's CPU-based implementation for the SSV kernel (stage).}
\label{fig:ssvcomparehmmer}
\end{figure*}

A comprehensive performance comparison is also made between optimized CUDAMPF++ and other implementations. Fig~\ref{fig:msvcomparehmmer} and~\ref{fig:ssvcomparehmmer} present results of comparison between CUDAMPF++ and CPU-based MSV/SSV stages with two datasets. The CUDAMPF++ achieves upto $282.6$ ($283.9$) and $465.7$ ($471.7$) GCUPS for MSV and SSV kernels, respectively, given the \textit{env\_nr} (\textit{est\_human}) dataset. Compared with the best performance achieved by dual Xeon E5-2650 CPUs, a maximum speedup of $168.3$x ($160.7$x) and a minimum speedup of $1.8$x ($1.7$x) are observed for the MSV (SSV) kernel of CUDAMPF++.

In the original HMMER3 paper~\cite{Eddy2011Accelerated}, Eddy reports $12$ GCUPS for MSV stage, achieved by a single CPU core. Several acceleration efforts exist and report higher performance: (a) an FPGA-based implementation~\cite{Abbas2010Accelerating} yields upto 81 GCUPS for MSV stage; (b) Lin~\cite{Lin2014Implementing} inherits and modifies a GPU-based implementation of HMMER2~\cite{Walters2009Evaluating} to accelerate MSV stage of HMMER3, which achieves upto 32.8 GCUPS on a Quadro K4000 GPU; (c)~\cite{Neto2015Acceleration} claims the first acceleration work on SSV stage of latest HMMER v3.1b2 and reports the maximum performance of 372.1 GCUPS on a GTX570 GPU. To sum up, as shown in Fig.~\ref{fig:msvcomparehmmer} and~\ref{fig:ssvcomparehmmer}, the proposed framework, CUDAMPF++, exceeds all existing work and exhibits strong consistency in performance regardless of either the model length or the amount of protein sequences.

\section{Related Work}
As one of the most popular tool for the analysis of homologous protein and nucleotide sequences, HMMER attracts many acceleration attempts. The previous version, HMMER2, is based on Viterbi algorithm that has proved to be the computational bottleneck. The initial effort of GPU-based implementation for HMMER2 is ClawHMMER~\cite{Horn2005ClawHMMER} which introduces a streaming version of Viterbi algorithm for GPUs. They also demonstrate the implementation running on a 16-node GPU cluster, each equipped with a Radeon 9800 Pro GPU. Another early GPU-based implementation is proposed by Walters~\textit{et al.}~\cite{Walters2009Evaluating} who properly fit the Viterbi algorithm into the CUDA-enabled GPU with several optimizations, like memory coalescing, proper kernel occupancy and shared/constant memory usage, which outperforms the ClawHMMER substantially. Yao~\textit{et al.}~\cite{Yao2010cuHMMER} present a CPU-GPU cooperative pattern to accelerate HMMER2. Ganesan~\textit{et al.}~\cite{Ganesan2010Accelerating} re-design the aligment process of a single sequence across multiple threads to partially break the sequential dependency in computation of Viterbi scores. This helps building a hybrid task and data-level parallelism that eliminates the overhead due to unbalanced sequence lengths.

However, with the heuristic pipeline, HMMER3 achieves about $100$x to $1000$x speedups over its predecessor~\cite{Eddy2011Accelerated}, which hence renders any acceleration effort of HMMER2 obsolete. There are only few existing work that aim to accelerate SSV, MSV and P7Viterbi stages of \textit{hmmsearch} pipeline in HMMER3. Abbas~\textit{et al.}~\cite{Abbas2010Accelerating} re-writes mathematical formulas of MSV and Viterbi algorithms to expose reduction and prefix scan computation patterns which are fitted into the FPGA architecture. In~\cite{li2010A}, a speculative method is proposed to reduce the number of global memory access on the GPU, which aims to accelerate the MSV stage. Lin~\textit{et al.}~\cite{Lin2014Implementing,Lin2015Accelerating} also focus on MSV stage but incorporate SIMD video instructions provied by the CUDA-enabled GPU into their method. Like the strategy of~\cite{Walters2009Evaluating}, they assign each thread to handle a whole sequence. A CPU-based implementation of P7Viterb stage is done by Ferreira~\textit{et al.}~\cite{Ferreira2014Cache} who propose a cache-oblivious parallel SIMD Viterbi algorithm that offsets cache miss penalties of original HMMER3 work. Neto~\textit{et al.}~\cite{Neto2015Acceleration} accelerate the SSV stage via a set of optimizations on the GPU, such as model tiling, outer loop unrolling, coalesced and vectorized memory access.

\section{Discussion}
While we have shown that the proposed framework with hierarchical parallelism achieves impressive performance based on Kepler architecture, we believe that more advanced GPU architectures, like Maxwell, Pascal and Volta, could also benefit from it because of its hardware-based design. It is easy to port the framework to run on advanced GPUs and gain better performance given more available hardware resources, such as on-chip registers, cache capacity, memory bandwidth and SMs. Also, the framework naturally has linear scalability when distributing protein sequences to multiple GPUs. To handle large models that exceed the carrying capability of single GPU, however, one potential solution is the model partitioning that distributes different segments of model to different GPUs while introducing inter-device communication (i.e., max-reduction, reordering). The multi-GPU implementation of the proposed framework is being investigated.

As for the general applicability, not only is the framework suitable for accelerating analogous algorithms of genomic sequence analysis, other domain-specified applications with some features may also benefit from it. The highlight features, for example, may include data irregularity, large-scaled working set and relatively complex logic with execution dependency. In the contrary, for some agent-based problems that usually investigate the behavior of millions of individuals, such as molecular dynamics or simulation of spatio-temporal dynamics, our framework may not be the preferred choice. Actually, the key performance factor is the innermost loop, corresponding to 3rd, 4th and 5th tiers of the proposed framework, in which we should only put necessary operations. In general, assuming that kernels are bound by the innermost loop, there are several suggestions related to minimizing the cost inside the innermost loop: (a) try to hold repeatedly used values in registers to avoid high-frequency communications between on and off-chip memory; (b) pre-load data needed by the innermost loop in outer loops; (c) use either L1 or texture cache to reduce the overhead of load/store operations; (d) try to use high-throughput arithmetic instructions. (e) use shuffle instructions rather than shared memory, if applicable.

Ultimately, this work sheds light a strategy to amplify the horsepower of individual GPU in an architecture-aware way while other acceleration efforts usually aim to exploit performance scaling with muliple GPUs.


\section{Conclusion}
In this paper, we propose a five-tiered parallel framework, CUDAMPF++, to accelerate computationally intensive tasks of the homologous protein sequence search with profile HMMs. This framework is based on CUDA-enabled GPUs, and it aims to fully utilize hardware resources of the GPU via exploiting finer-grained parallelism (multi-sequence alignment) compared with its predecessor. In addition, we introduce a novel idea that improves the performance and scalability of the proposed framework by sacrificing L1 CHR proactively. As shown by experimental results, the optimized framework outperforms all existing work, and it exhibits good consistency in performance regardless of the variation of query models or protein sequence datasets. For MSV (SSV) kernels, the peak performance of the CUDAMPF++ is $283.9$ ($471.7$) GCUPS on single K40 GPU, and impressive speedups ranging from $1.8$x ($1.7$x) to $168.3$x ($160.7$x) are achieved over the CPU-based implementation (16 cores, 32 threads). Moreover, further generalization of the proposed framework is also discussed.


%



\ifCLASSOPTIONcompsoc
  \section*{Acknowledgments}
\else
  \section*{Acknowledgment}
\fi

The authors would like to thank the NVIDIA-Professor partnership for generous donations in carrying out this research.

\ifCLASSOPTIONcaptionsoff
  \newpage
\fi



\bibliographystyle{IEEEtran}
\bibliography{bare_jrnl_compsoc}





%

\begin{IEEEbiography}{Hanyu Jiang}
Biography text here.
\end{IEEEbiography}

\begin{IEEEbiography}{Narayan Ganesan}
Biography text here.
\end{IEEEbiography}

\begin{IEEEbiography}{Yu-Dong Yao}
Biography text here.
\end{IEEEbiography}







\end{document}